\documentstyle[12pt,aaspp4]{article}

\lefthead{Noguchi}
\righthead{Early Evolution of Disk Galaxies}
\begin{document}

\title{Early Evolution of Disk Galaxies : Formation of Bulges \\
    in Clumpy Young Galactic Disks}

\author{Masafumi Noguchi}

\affil{Astronomical Institute, Tohoku University,
Aoba-ku, Sendai 980-77, Japan;
noguchi@astroa.astr.tohoku.ac.jp}

\begin{abstract}

A new idea is proposed for the origin of bulges in spiral galaxies.
Numerical simulations for the protogalactic collapse suggest strongly
that galactic bulges have been assembled from massive 
clumps formed in the galactic disks in their early evolutionary phase.
These clumps result from the gravitational instability of the  gas-rich 
disks of young galaxies.
Those massive clumps, individual masses of which can be 
as large as $\sim 10^9 M_{\odot}$,
 are able to spiral toward the galactic center
owing to dynamical frictions within a few Gyr.
Inward transport of disk matter by this process leads to the formation
of a galactic bulge. 
A simple analytical model has been constructed, in which the clumpy evolution
of a disk galaxy is controlled by two parameters; the
timescale with which the primordial gas in the halo accretes onto the disk plane
(i.e., collapse timescale) and the mass fraction of the primordial gas relative to the galaxy total mass.
Under plausible assumptions for the variation of these parameters among 
spiral galaxies,
the clumpy evolution model
can explain the observed trend that the bulge-to-disk ratio increases as the total mass
or the internal density of the galaxy increases.
This success suggests that the clumpy evolution of the galactic disk constitutes
an important ingredient of the disk galaxy evolution.
Star formation in primeval disk galaxies takes place mostly in the clumps. Resulting 
knotty appearance of these systems may explain the peculiar morphology observed in a number
of high-redshift galaxies.

\end{abstract}


\section{Introduction}

One of the most notable structural features of disk galaxies is that they are
composed of two distinct components; disks and bulges.
The origin of this two-component structure remains unclear, although it constitutes
an important  back-bone of disk galaxies.
Several theoretical studies have addressed this problem (e.g., \cite{lar76,got76}).
Conventional understanding is that the galactic bulges have been formed as a single body
within a relatively short period, as a result of the collapse of a gaseous
protogalaxy, and the later accretion of the residual
primordial gas has built the disk components.
Pioneering models by Larson (1976) for the formation of disk galaxies
are one of the most extensive numerical models which have embodied this idea, but
have a potential problem that two different regimes of star formation should be 
assumed in order to realize distinct separation between a bulge and a disk.
Recent observations of the Milky Way bulge (e.g., \cite{ric96,mcw94}) and bulges of
other galaxies have stimulated attempts to elucidate when and how the galactic bulges 
have been formed and nature of the bulges itself (e.g.,\cite{mat90,pel96,kor93}).

On the other hand,
recent observations by the Hubble Space Telescope (HST) and other new
generation ground-based telescopes are revealing
the properties of distant galaxies at the epochs when the Universe was
younger than half of its present age, and thus providing a direct access 
to the initial evolution phase of galaxies. Those galaxies having a redshift
$z> \sim 1$ generally exhibit morphologically peculiar structures which defy 
the application of the traditional Hubble classification scheme
(e.g., \cite{van96,abr96,gla94,gla95,gri94,dri95,cow95}).
The most remarkable class of those high-redshift galaxies includes the systems which 
consist of a few blobs (clumps) embedded in a common diffuse envelope.
For example, Griffiths et al. (1994) note several galaxies with prominent bright knots
in their sample of HST Medium Deep Survey. Galaxies having similar morphology are
also found in images examined by other groups (e.g., \cite{gia96b,koo96,van96}).

Some of the peculiarities observed in these high-redshift galaxies may originate in
copious interstellar gas existing in early evolution phase of disk galaxies.
In most disk galaxies which we observe in the present-day universe, the interstellar gas
can be regarded as a {\it dynamically} passive component because of the small fraction
it occupies in the total mass of the galaxy. The motion of the gas is determined practically
by the gravitational field arising from the stellar component, and apart from
dissipative process and star formation, ballistic motion is a good approximation 
for the description
of the gas kinematics. This situation, however, changes drastically 
if the mass fraction of the gas relative to the total galaxy mass exceeds
about 10 percent. 
In this gas-rich regime, the interstellar gas is very active {\it dynamically}.
The gas, being dissipative and self-gravitating, tends to form numerous massive clumps, each of which is gravitationaly
bound. These massive clumps are involved with various dynamical processes (\cite{nog98}).
For example, they deflect stellar motions effectively and heat up the stellar disk component
(i.e., increase the stellar velocity dispersion), and thus affect the stability of the 
galactic disk
(\cite{shl93,nog96}).
Noguchi (1998) has proposed the gravitational instability origin for those sub-galactic clumps
frequently observed in high-redshift galaxies,
and argued that they may serve as building blocks of galactic bulges. The present paper expands upon
this idea and tries to discuss the evolution
of young disk galaxies in more details, with much emphasis placed on the formation
 of galactic bulges and dependence of its efficiency on galaxy parameters.

\S2 and \S3 describe the numerical simulation which has inspired the new idea of
bulge formation.
In \S4, an analytical modelling is devised for the growth of galactic disks
and is applied to bulge formation.
Observational data about global properties of spiral galaxies are summarized in \S5,
and compared with the theoretical results in \S6. Discussion and conclusions are 
given in \S7 and \S8, respectively.

\section{Numerical Simulation}

This section provides a full account of the numerical model which has motivated
the new idea of bulge formation briefly given in Noguchi (1998) and examined 
extensively here.
Collapse of a protogalaxy composed of the dark matter and primordial gas has been
simulated by an N-body method including a star formation algorithm. Similar approaches
have been taken by many researchers (e.g.,
\cite{bar87,kat91,kat92,ste95,nav94}).
Unlike most of these works, which employed initial conditions based on a specific type
of cosmology such as the Cold Dark Matter model,
the present work 
 does not assume a particular cosmology.
I start from an idealized initial condition and attempt 
to predict properties of young disk galaxies which will be robust to the nature of the 
assumed cosmology.
The primary interest here is in investigating detailed morphological 
and dynamical evolution of forming disk galaxies. Therefore the simulation is restricted 
to a three dimensional volume which encloses a single galaxy.
The factor by which a galaxy has collapsed after turn-around (i.e., separation
from Hubble expansion) is considered to be large, a factor of 10 (e.g., \cite{fal80}).
The present study does not deal with the whole process of this collapse but only the evolution
of the galaxy after it has shrinked to the present size and has become balanced 
by the increased centrifugal force in the direction perpendicular to the rotation
axis. Therefore, the halo (the dark matter) mentioned hereafter means only its portion
contained in the optical radius of the galaxy.
This limitation in space makes it possible to investigate more detailed structure
and kinematics of the forming galaxy than other cosmologically-implemented
simulations.

The primordial gas which has been destined to make the visible parts of the
present-day galaxies is considered to have been more-or-less clumpy
(e.g., \cite{fal85}). Sizes and masses of these clumps are matter of
debate. Though the recent progress in the observational technique is enabling
access to these pregalactic entities, in the form of quasar absorption line systems for example,
the physical properties deduced from those observations are diverse.
For example, Steidel (1990) estimates the typical size of 1 - 15 kpc
and the mass of $10^6 - 10^9 M_{\odot}$ for the Lyman limit systems, which are
considered to constitute halos of galaxies located on the line-of-sights
to more distant quasars.                                      

Depending on the assumed size and mass of these building blocks, two extreme
pictures can be considered to describe the formation process of  galaxies
in general. The first one
envisions a relatively regular and smooth (both spatially and temporarily)
gas accretion process in the potential well of the protogalaxy (e.g.,
\cite{got76,lar76}). Graininess of the
pregalactic clumps is neglected on the scale of galaxy size. 
The second picture, in another extreme, considers that a typical galaxy 
has been made by mergers of subgalactic units, which are not so small and numerous
as to permit neglect of their discreteness (e.g., \cite{col94}).

Although the evolution of elliptical galaxies could be much different
depending on which picture is adopted, these two limiting scenarios seem 
to converge in the case of disk galaxies. Stellar dynamical processes, which retain memory
of initial conditions to some extent, are considered to have non-negligible effects
on the formation and evolution of elliptical galaxies. In contrast with this,
the observed thiness of the galactic 
disks suggests that the assembling of pregalactic units into galactic disks
has been a highly dissipative process. Merger of non-dissipative stellar systems
would puff up the disk (e.g., \cite{tot92}). In other words, pregalactic bodies,
whatever their sizes and masses may have been, should have been primarily gaseous when 
they coagulated to make a galactic disk. Dissipative nature of these bodies must have
erased identity of individual bodies quickly. Thus, all the disk galaxies, in their
early evolutionary phase, should have had
a relatively smooth and mostly gaseous galactic disk with roughly
the same radius and thickness as the present-day stellar disk,
both in the smooth collapse picture and in the merger scenario. 

\subsection{Initial Conditions}

As a device to realize dissipative growth of a galactic disk,
we consider a spherical protogalaxy consisting of dark matter and primordial gas.
These two components have  uniform density distributions with the same cut-off
radius, $R=15 kpc$. The total
mass of the system is $M=1.5 \times 10^{11}M_{\odot}$, with the
masses of the dark matter and gas components being $M_h=0.75 \times 10^{11}M_{\odot}$ and
$M_g=0.75 \times 10^{11}M_{\odot}$, respectively.
With this choice, the dynamical timescale of the system, $(R^3/GM)^{1/2} = 7.06 \times 10^7yr$,
where $G$ is the gravitational constant.
The unit velocity becomes $(GM/R)^{1/2}=208 kms^{-1}$.
The dark matter and gas components are modeled as systems comprising $N_h=20000$ and
$N_g=30000$ particles, respectively.
The adopted gas fraction, $M_g/M=$ 0.5, is based on the observations that
the mass fraction of the dark halo component {\it within the optical radius} is
clustering around 0.5 for most disk galaxies for which the rotation curve and the
distribution of the luminous component are observed with high accuracy (\cite{van82,bah85}, but see
also discussion in \S4.6).
There is no stellar particle initially ($N_s$=0).

The dark matter particles are given isotropic random motions with an one-dimensional
velocity dispersion of $\sigma=122 kms^{-1}$, bringing the dark halo into
virial equilibrium roughly. The gas particles are given a uniform rotation with an
angular frequency of $\Omega=13.9 kms^{-1}kpc^{-1}$. 
Therefore, the gas system is
 already in rough centrifugal equilibrium owing to enough rotational support, and
is expected to collapse not much in the radial direction but mostly along the spin axis.
The direction of the angular momentum vector for
this rotational motion is taken as the $z$-axis hereafter, with the $x$- and $y$- axes
lying in the plane perpendicular to the $z$-axis.
The evolution of the system under its self-gravity is simulated by an N-body method.

The dark matter particles are treated as collisionless. Namely, their motions are 
determined solely by the global gravitational field they feel. On the other hand, gas particles
are modeled as particles which collide with each other inelastically, thus dissipating
their kinetic energy. Star formation process from the gas is also included as described shortly.
The gravitational force exerted on each particle is calculated by the GRAPE, which is a
dedicated device for N-body calculations (\cite{ebi93}).  
The softening radii, $\varepsilon_h=300pc$ and $\varepsilon_g=150pc$, are introduced for dark matter
and gas particles, respectively, to suppress undesirable two-body effects.

\subsection{Gas Dynamics and Star Formation}

Gas dynamics is treated by the so-called sticky particle method.
Inelastic collisions between the gas particles are introduced
to model the dissipative nature of the interstellar medium
(\cite{lev81,rob84,hau84}).
All the cloud particles are assumed to have the same  finite radius $r_c=38pc$.
In the simulation, two overlapping clouds are made to collide inelastically 
provided that they are approaching each other.
After collision, the radial component of the mutual velocity is multiplied by 
$f_{col}=0.01$ and its 	sign is reversed, while
the tangential component is unchanged, in order to mimick energy dissipation.
The evolution of the system does not change appreciably  as far as  $f_{col}< \sim 0.5$.

The star formation process is simulated by changing a gas cloud particle to a 
stellar particle.
This conversion is performed with a probability which is related to the local
gas density around the cloud particle
as follows. The local gas density, $\rho$, for a given
cloud particle is determined by counting the number
of the gas clouds residing in the sphere of a radius $r_{dens}=750pc$ centered on the cloud
in question. Then the probability, $p$, for star formation for this cloud
in the current time step is calculated by,
$p = K_{star} \times dt \times \rho ^{1/2}$.
Here, $K_{star}$ is a coefficient which controls the star formation efficiency, and
$dt$ is the size of the time step, which is $7.06 \times 10^5 yr$ (i.e., one hundredth of the
dynamical time).
The value of $K_{star}$ is determined empirically so that the time variation of star formation
rate is in a rough agreement with the one inferred for disk galaxies from
various observational data
(see \S3).
The square root dependence on the local gas density expressed in the above expression
for $p$ is 
specified by the consideration that the star formation process, though its detailed
nature is not well understood, originates in some form of local gravitational instability
in the interstellar medium, and hence its characteristic timescale must be
 related to the free-fall
timescale of the local unstable region. 

After getting $p$ by the above equation, one number $\xi$ is created
in the range (0,1) using the uniform random number generator.
If $\xi < p$, that cloud particle is changed into a stellar one. 
If $\xi > p$, no such transformation is made.
This recipe for star formation, namely converting a whole cloud into a stellar 
particle
with some probability instead of converting some fraction ($p$ in this case) of one cloud
at
every time step, was taken to avoid intolerable increase of the number
of particles in the model as new stars form. 
This procedure makes the sum of the number of gas particles, $N_g$, and that of
stellar particles, $N_s$, to be constant during the course of simulation.
One important addition to this recipe is specification of a threshold density
for star formation. Star formation is inhibited if the local gas density is less than
a critical value $\rho_{th}$ (see below). 
Stellar particles are treated as collisionless, with a 
gravitational softening parameter, $\varepsilon_{s}=150pc$.

The threshold density was set to be $\rho_{th}=0.1$ ($M_{\odot} pc^{-3}$).
Observations of nearby disk galaxies suggest that the 
threshold {\it surface} gas density, $\mu_{th}$, for star formation is $\sim (1-10) M_\odot pc^{-2}$
(\cite{ken89}).  Because the gas in the present model occupies a three-dimensional
volume in general and is not necessarily confined in a planar configuration, the application 
of a surface density sounds questionable and a threshold {\it volume} density will be more
relevant. Using the observed thickness, $h_g \sim$ 100 pc,
 of the distribution of the interstellar molecular gas in the disk of
the Milky Way galaxy (e.g., \cite{sol79}), I converted the 
threshold surface density to the 
threshold volume
density like $\rho_{th} \sim \mu_{th}/h_g \sim  
(0.01-0.1)M_{\odot} pc^{-3}$.
The value adopted for $\rho_{th}$ in the present simulation is the upper limit of this range, though
such a translation from the two-dimensional density to a three-dimensional one may not
be justified fully, because the configuration of the interstellar matter could affect
its stability and ability to form stars.

\section{Numerical Results}

Morphological evolution of this model is shown in Figure 1.
As a galactic disk builds up by infall of the gas from the halo region, several massive clumps
are formed in the disk plane [$\sim 0.6 <t< \sim 0.7$, especially Figure 1(e)]. 
These clumps follow the global rotation of the galactic disk.
Some of these clumps are observed to merge and make larger
clumps while orbiting in the disk, as clearly seen in Figure 1(e). Because these clumps suffer from dynamical friction
against surrounding disk material, they gradually sink to the central region of the disk.
This transfer of matter 
leads to an accumulation of a large mass in the central region.
The side view of the system is quite intriguing [Figure 1(f)].
It is noted that this inward motion of clumps leads to the formation of a spheroidal component
in the galactic center. It is natural to regard this component as a galactic bulge.
Note that almost all the star formation takes place in the disk plane [Figure 1(d)].
The bulge component is not made by early star formation occuring in a
spheroidal space at the galactic center.

The clumpiness of the system makes definition of the galactic center not a trivial problem.
In the analyses described below, the galactic center is defined to be the center of the most massive
clump at each epoch, which is called the bulge. The galactic center thus defined can be displaced from the 
coordinate origin, $x=y=z=0$. The mean disk plane, defined as the peak position in the z-distribution
of the gas clouds, also drifts slightly in the negative z-direction during the simulation. 
These deviations from the coordinate system are taken into account in the analysis when
appropriate.

Specification of a threshold gas density for star formation is definitely responsible 
for the lack of star formation before the epoch of disk formation. Volume density of the gas
is very low when the gas occupies a three-dimensional spherical volume of the halo region,
but increases by a large amount when the gas is gathered into a two-dimensional disk configuration. 
This change of the dimensionality initiates vigorous star formation in the galactic disk.
On the other hand, the column density of the gas hardly changes because the collapse occurs
mostly in the vertical direction.

Clumpiness observed in the numerical model is a direct consequence of the gas-richness
of the disk which forms by a relatively rapid infall.
Time evolution of the masses of the total disk and the gas disk is shown in Figure 2.
Here, 
the gas particles whose distance from the mean disk plane is less than 0.1
$R$ are considered to constitute the gas disk,
and all the stellar particles
 are regarded as members of the stellar disk.
The regions with the cylindrical radius $r<0.3 R$ or $r>1.5 R$ are excluded.
Initial evolution is the growth of a mostly gaseous galactic disk
along the line $f_g=f_d$. After the disk formation has been almost
completed, star formation 
starts to deplete the gas component, with the model point going down 
nearly along the line, $f_d$= constant, in the left panel.
A slight decrase in $f_d$ is due to inflow of the clumps toward the galactic center, which builds up the
bulge component.

The clumps in this numerical
model start to appear distinctly at $t \sim 0.6$ Gyr, when the value of $f_g$
is around its maximum of 0.19.
I selected a few typical clumps at $t=0.64$ Gyr as shown in
Figure 1(e) and investigated their internal structure.
Examination of the surface density profile made it possible to draw a boundary
between the main body of the clump and its envelope and to calculate the mass of the main body 
as the clump mass.
Masses of individual clumps are found to be of the order of 0.01$M$ ($\sim 10^{9} M_\odot$).
This value far exceeds
those of any molecular complexes observed in nearby galaxies
(i.e., the most massive known entities
which constitute the galactic disks in the present epoch),
and can be well compared to the masses of dwarf galaxies. 

The dashed line in Figure 2 indicates the evolutionary track of an analytical model which is described
later. This analytical model, with a collapse timescale of 0.3 Gyr and an initial
fraction of the gas mass of 0.5, represents the behavior 
of the numerical model fairly well.
I calculated
the expected clump mass as the instantaneous Jeans mass in the growing model disk 
of this analytical
model (open circles in Figure 2, right panel).
 The clump mass thus obtained reaches the maximum value
of $\sim 10 ^9 M_{\odot}$ at $t \sim 0.5$ Gyr when $f_g$ (dashed line)
reaches the maximum of $\sim 0.14$, which 
is roughly equal to the maximum of $f_g$ in the numerical model.
The agreement between the numerical and analytical clump masses suggests that the
seeds of those clumps in the numerical model
are created by the local gravitational instablility in the 
gaseous disk component.
We should, however, be careful in relating the clump mass to the Jeans mass in the
 disk at current epoch,
because masses of individual clumps in the numerical model
increase continuously in general due to
mergers among themselves and later accretion of the surrounding gas,
whereas the analytically calculated clump mass starts to decrease as $f_g$ decreases.

Figure 3 shows the time development of the star formation rate (SFR).
The SFR starts to increase abruptly $\sim 0.5$ Gyr
 after the start of the simulation,
and attains a peak rate of $\sim 40 M_\odot yr^{-1}$ at $t \sim  0.7$ Gyr. 
This is just the epoch when most of the gas has settled to the disk [see Figure 1(b)].
The SFR declines after the maximum monotonically except a few weak but distinct burst-like increases. 
By $t=2.1$ Gyr, 37 percent of the
gas has been converted into stars.
The peak SFR of $\sim 40 M_\odot yr^{-1}$ is within the range of 4-75 $M_{\odot} yr^{-1}$
deduced by Steidel et al. (1996b) for a population of $z>3$ galaxies.
The temporal change of the SFR in the numerical model agrees also with the  history of star formation
in early-type disk galaxies inferred from the observations (\cite{san86}).
Note also that the overall behavior of the SFR is similar to that of the SFR obtained 
in the analytical model.

It is noted that the SFR reaches the maximum roughly at the time when the clumpy
structure is most prominent (Figure 3). This means that, during the course of evolution, the
model galaxy will be observed as a clumpy object at the highest probability, provided the observation
is made in the rest-frame $UV$ band, which is sensitive to young massive stars.
Also note global asymmetry of the stellar component
in the present model, which is most noticeable at $t$= 0.9 Gyr to 1.3 Gyr
 [see Figure 1(e)].

The asymmetric and clumpy disk seen in this  model
 is strongly reminiscent of the images of
many galaxies at large redshift obtained by the HST (e.g., \cite{van96,gri94,koo96}).
Although the irregular and
clumpy nature is most noticeably found in the objects with a medium redshift (i.e.,
$ \sim 0.5 < z < \sim 1$), it may be traced into a larger redshift.
Steidel et al. (1996a) investigated the morphology of Lyman break galaxies
with $2.3 < z < 3.4$ found in the Hubble Deep Field and noticed some objects
having multicomponent structure.
They also mention diffuse asymmetric "halos" around these high-redshift galaxies.
In view of the present numerical result, 
some parts of these morphological peculiarities may be caused by gravitational
instability of the galactic disks in their early evolution phases.

The clumps discussed above experience strong dynamical friction owing to 
their large masses. Resulting accumulation of clumps to the galactic center
makes a bulge as already stated. 
The bulge in the present numerical model is not formed quickly in a single
 event as the protogalaxy collapses to the galactic center but
is assembled gradually as individual clumps formed in the 
disk plane spiral into the galactic center.
The growth of the bulge is stepwise because accretion of clumps
takes place in a discrete manner.
Figure 4 shows the distribution of the age for the bulge stars at several epochs.
Discrete nature of the bulge growth is manifested as a few local maxima in the 
age distribution (see the bottom panel, especially),
which correspond respectively to starbursts triggered by the falling-in of a massive clump into
the bulge region.
At the final epoch indicated ($t=1.98$ Gyr),
the age spread is as large as $\sim 1$ Gyr.

\section{Analytical Multi-Component Models}

The numerical model discussed above is enlighting, but has a difficulty.
As shown below, one of the most important parameters which govern the evolution
of disk galaxies is the accretion timescale, namely, the timescale of gas infall from the halo.
It is difficult to vary this timescale freely
by employing the sticky-particle method.
Accretion can be accelerated or slowed down
by varying the size of the clouds and/or the restitution
coefficient $f_{col}$ in the cloud collisions in the sticky-particle model.
However, it is not fully clear how these parameters
are related to the observed
properties of the interstellar gas. 
For example, the giant molecular clouds of the Milky Way galaxy have a very complicated
internal structure comprising  many hierarchical levels in spatial scale, and no single
spatial scale seems to dominate (e.g., \cite{hen91}).
Also, the accretion of gas to the disk plane is likely to be
affected by other effects than dissipative cloud-cloud collisions as discussed below.
These considerations have motivated development of  analytical models
which introduce the accretion timescale
as a free parameter.
In this section, I discuss the analytical model of disk growth to remedy the limitation
possessed by the numerical simulation.
The analytical model developed here has essentially the same ingredients as the 
previous version described in Noguchi (1996), but is improved in several respects.

\subsection{Basic Equations} 

I consider a spherical halo with a radius $R$. The primordial gas
initially distributed in this halo region gradually accretes to the disk plane and builds up
a galactic disk. As the accretion proceeds, the disk becomes to contain more and more gas (i.e.,
the interstellar gas) and stars form from it.
I assume that the stars and the gas in the disk occupy a flat cylindrical
region with the same radius as the halo.
The whole system considered here is meant to
represent the portion of a disk galaxy within its optical radius
so that the halo in the present
model should be regarded as not representing the entire dark halo (the massive halo) but its
portion within the optical extent of the galaxy, in accordance with the numerical model
presented above.
The system is divided into 
four components; the halo, the stellar disk, the gas disk, and the bulge.
The mass of each component is denoted by $m_h$, $m_s$, $m_g$, and $m_b$, respectively.
Total mass is denoted by $M$ (=$m_h+m_s+m_g+m_b$).
I am not concerned with any
structures which might develop in each component but treat each component as a single zone. 
Namely, the physical state of each component 
at a given time  is specified
by several global
quantities whose value should be understood as  a characteristic one, namely the average
over the entire region of that component.

Under this simplification,
time evolution of masses in respective components 
is formulated as follows.

$${ {dm_{g}} \over {dt}  }=- { SFR }
   -{ {m_g} \over {\tau_{fri}}}
  + {{M \Gamma t} \over {\beta^2}}  exp( -{ {t} \over {\beta}})
  , \eqno(1)  $$

$${ {dm_{s}} \over {dt}  }= { SFR }, \eqno(2) $$

\noindent{and}

$${ {dm_{b}} \over {dt}  }=
   { {m_g} \over {\tau_{fri}}}, \eqno(3)  $$

\noindent{where $t$ is the time} reckoned from the beginning of gas infall (which is here assumed to be 12 Gyr ago).
The first term in the righthand side of
equations (1) and (2) denotes the effect of star formation and $SFR$
here stands for the star formation rate for the whole galaxy.
$\tau_{fri}$ is the timescale of the inward motion of clumps formed in the disk.
This inflow is caused by dynamical frictions acting upon the clumps, and
its timescale determines the growth rate of the bulge component.
The bulge growth is determined by eq(3) in this analytical formulation.
The third term in the righthand side of equation (1) represents addition of 
gas to the disk by the accretion (gas infall)
from outside the disk plane, the timescale of which is denoted by $\beta$.
Hereafter, $\beta$ is referred to interchangeably as the accretion timescale,
the collapse timescale, or the infall timescale.
As the infall proceeds,  the mass of the halo decreases correspondingly, because
the total mass of the system, $M$, is assumed to be conserved
(Here the halo serves only as a reservoir of the primordial gas and its evolution
is not traced).
The total mass of the matter 
which eventually accretes to the disk is specified by its fraction, $\Gamma$,
with respect to the total mass of the galaxy.
Namely, $\Gamma$ is the fraction of the gas mass relative to the total galaxy
mass at the initial instant.
The functional form of the time variation of 
the infall rate in eqation (1) was adopted because  
it is physically reasonable and mathematically convenient.
The time dependence here is slightly modified from the previous one adopted in Noguchi (1996).

\subsection{Star Formation Rate}

The fundamental process of star formation in the interstellar medium is not yet
well known. Phenomenologically, the star formation rate is often represented as a power
of the amount of interstellar gas (Schmidt law). 
The present study follows the result by Kennicutt (1989), who has found from a compilation
of data for a number of nearby disk galaxies that the star formation rate per unit area
of the galactic disk is proportional to $N$-th power of the surface density of the gas
which includes both molecular and atomic hydrogen.
In the present model, this relation is represented as,

$$ SFR = \alpha  \Sigma_g^{N} R^2  ,   \eqno{(4)} $$

\noindent{where $\Sigma_g$ denotes the gas surface density in the disk and
is approximated by $m_g/(\pi R^2)$ in the present multi-component
treatment, and $\alpha$ is a coefficient which determines
the absolute value of the star formation rate.}
The range of the power in the above Schmidt law is $N = 1.3 \pm 0.3$ according to
Kennicutt (1989). 
It was found that setting $\alpha=355 M_{\odot} yr^{-1}$ produces a range of star formation rate
compatible with the observational inference,
when the mass, $m_g$, and the radius, $R$, are expressed in units of $10^{11} M_{\odot}$ and 10 kpc,
respectively.
Therefore $\alpha$ is fixed to $355 M_{\odot} yr^{-1}$
and $N$=1, hereafter.
Adoption of an universal value of $\alpha$ for all the models is based on the inference
that the fundamental star formation process is the same in all the disk galaxies.

The work by Kennicutt (1989) suggests that a certain threshold exists for gas density,
below which star formation is effectively inhibited. One plausible interpretation is that the star
formation activity is associated closely with the gravitational instability
 of the gas disk and the threshold 
corresponds to neutral stability. How is the threshold density determined? One possibility is that
the galactic disks have
 an intrinsic lower limit of velocity dispersion in the gas cloud motion, and a
disk with a too small surface density is stabilized by the random motion of the interstellar 
gas  clouds. Indeed, HI and CO observations of nearby galaxies and the 
Milky Way galaxy suggest near constancy 
among different galaxies of the 
velocity dispersion in the atomic and molecular gas clouds (e.g.,
\cite{van84,sta89}).
Therefore, I introduced a threshold for star formation by specifying a minimum velocity 
dispersion which the gas component of the galactic disk can take.

The star formation threshold introduced in the present model works as follows.
As explained shortly, the gas disk is assumed to stay at the marginal stability defined by $Q=1$,
where $Q$ denotes the stability parameter introduced by Toomre (1964). At this state,
the surface density of the gas and the velocity dispersion (or the sound velocity), $\sigma$, in the gas are
related by $\pi G \Sigma_g = \sigma \kappa$, where $\kappa$ is the epicyclic frequency.
As $\Sigma_g$ decreases in late phase of evolution, $\sigma$ also decreases correspondingly.
However, $\sigma$ cannot become less than $\sigma_{min}$, which is the specified lower limit of the velocity
dispersion. Therefore, star formation stops at the instant when $\sigma$ reaches $\sigma_{min}$. 
The threshold gas surface density is then determined by $\Sigma_{min}=\sigma_{min}\kappa/\pi G$.
After this instant, the amount of the gas consumed by star formation is balanced with
the amount of the gas added to the disk by accretion, so that the gas surface density
is always kept nearly at the threshold value (i.e., $\Sigma_g \sim \Sigma_{min}$).
As a special case, setting $\sigma_{min}=0$ allows star formation to proceed continuously
depending on the current gas density.

\subsection{Dynamical Friction Timescale}

Noguchi (1996) applied the classical Chandrasekhar formula in evaluating $\tau_{fri}$.
It is, however, doubtful that this formula, which assumes homogeneous and infinite background of
randomly moving
light particles, can be safely applied to a heavy body orbiting in a highly flattened
and systematically rotating disk component.
Quinn \& Goodman (1986) have extensively discussed sinking process of satellite galaxies
through galactic disks and suggest several analytical evaluations of the orbital decay
timescale.

Instead of applying a certain analytical formula, I have resorted to a more empirical approach in evaluation 
of $\tau_{fri}$.
I have run a number of numerical simulations for a massive rigid body (the clump,
hereafter) orbiting in the disk plane of a 
galaxy model, varying the clump mass, the disk surface density, the velocity
dispersion of disk stars, and the shape of the rotation curve (see Appendix for numerical details).
Time variation of the galactocentric distance of the clump in each model is plotted in Figure 5.
Because all the models show a remarkably similar behavior when the time axis is 
suitably adjusted, I 
decided to define $\tau_{fri}$ as the time it takes the clump to move from the initial radius,
$r=0.8$,  to  the final radius, $r=0.2$.

Measurement of $\tau_{fri}$ in these experiments is summarized in Table 1. It is evident that
the mass of the clump and the surface density of the disk are the key parameters which
determine the dynamical friction timescale. Dependence of $\tau_{fri}$ on other parameters
is weak and can be neglected for the practical range of interest. 
This is fortunate because little is known in this analytical treatment about
the time variation of the disk velocity dispersion or the shape of
the rotation curve. Examination of the parameter dependence
has led to the following empirical formula for $\tau_{fri}$.

$$ {{\tau_{fri}} \over {\tau_{dyn}}}= 0.25 ({{m_{cl}} \over {M}})^{-0.5}
( {{\Sigma} \over {M/R^2}}) ^ {-0.67}.  \eqno(5) $$

\noindent{Here, $\tau_{dyn} [\equiv (GM/R^3)^{-1/2}]$ is the dynamical time
of the galaxy, $m_{cl}$ is the mass of the clump, and $\Sigma$ is the surface density 
of the total disk which includes both gas and stars, i.e., 
$\Sigma=(m_g+m_s)/(\pi R^2)$}.

In calculating the clump mass, it is assumed that
 the gas disk is always maintained in the marginally unstable state with
$Q$=1. 
This assumption finds justification as follows. If $Q>1$ at a certain moment, no star formation
is expected to occur, because the gas disk is stable gravitationally to small scale
perturbations. Then $Q$ decreases, because  heat input from massive stars is lacking and the gas 
radiates its energy. When the decreasing $Q$ cuts the value of unity
and becomes slightly smaller than unity, the instability sets in and stars begin to form.
These stars provide the gas with energy through supernova explosions and stellar winds.
Then $Q$ is elevated above unity again. Thus the value of $Q$ will oscillate around $Q=1$,
and $Q=1$ is a good approximation of the dynamical state of the gas disk.

In the state of marginal instability (or stability), the mass of the clump formed is
given by,

$$ m_{cl} = \pi(0.5\lambda_c)^2\Sigma_g =
  {{\pi^5 \Sigma_g^3} \over {\kappa^4}},     \eqno(6)  $$

\noindent{where} the critical wavelength
$  \lambda_c = {{2\pi^2\Sigma_g} \over {\kappa^2}}$ and
the epicyclic frequency is approximated as
$ \kappa=(2M/R^3)^{0.5}$.
Clump formation is related to star formation closely in the present model.
To be consistent with introduction of a star formation threshold, clump formation is inhibited
when $\Sigma_g < \Sigma_{min}$.
Instantaneous value of $\tau_{fri}$ is determined by combining equations (5) and (6).
The bulge mass, $m_b$, thus calculated through equation (3)
may be uncertain by a considerable amount because of
many simplifications made in the present formulation.
However,  relative comparison
of $m_b$ between different models will be meaningful, and I focus on qualitative 
behavior of the bulge growth in what follows.
 
Now equations (1), (2), and (3), combined with auxiliary equations
(4), (5), and (6), complete a set of  equations which 
determine the temporal evolution of the system. 
By integrating these equations numerically starting from the initial condition,
$m_g=m_s=m_b=0$, we know time evolution of any physical quantities of
interest.

\subsection{Important Parameters: $\beta$ and $\Gamma$}

Evolution of a galactic disk in the present formulation
is determined mostly by two parameters, the accretion
timescale, $\beta$, and the  mass fraction of the primordial gas, $\Gamma$.
Before dealing with more complicated cases, I here discuss what effect these parameters
have on the disk evolution. 
Two model series are considered,
in order to isolate the effect of varying each parameter.
In Series A, only the value of $\beta$ is changed with
all other model parameters being equal, while all the models in Series B have the same parameters
but $\Gamma$.
The mass and the radius of the galaxy are fixed to $M=10^{11} M_{\odot}$ and $R=10 kpc$
in both series.

Figure 6 shows time evolution of the models in Series A.
It is seen that, as $\beta$ decreases, the peak values of the gas mass fraction ($m_g/M$), the star formation
rate, and the clump mass become larger and are attained at a progressively earlier epoch.
On the other hand, the present-day values of the gas mass fraction,
the star formation rate, and the clump mass
are larger for a model with slower accretion.
The bulge formation starts and finishes at an earlier epoch, as $\beta$ decreases.
A smaller $\beta$ also leads to a larger bulge. The mass of the stellar disk at the present
epoch is smaller for a smaller $\beta$, because a larger fraction of the accreted material 
goes to the bulge component in this case, due to more efficient inflow of matter
caused by more massive gas clumps.
Thus the models in Series A show qualitatively different time evolution depending upon the value
of $\beta$.
It should be noted that the bulge growth caused by the inflow is heavily reduced
or stopped when the disk becomes mostly stellar.

What about the effect of varying $\Gamma$? Figure 7, which plots temporal behavoir of Series B,
indicates that all the models in this family exhibit a qualitatively similar time evolution.
Each quantity changes as time in a similar way, and no remarkable effect of varying 
$\Gamma$ is observed. Effect of varying $\Gamma$ manifests only as the difference in the absolute
value.
Dotted lines in Figure 7 indicate the values divided by $\Gamma$ of each model. Rough equality
in the normalized values of $m_s/M$, $m_g/M$, and SFR among different models indicates that
these quantities are roughly proportional to the mass fraction, $\Gamma$.
On the other hand, the normalized clump mass and the normalized bulge mass show
a difference of $\sim 10$ among the calculated models. Thus these quantities behave in a highly
nonlinear way with respect to $\Gamma$. This strong nonlinearity leads to a large
difference in the ratio of the bulge to the disk, or the $B/T$.

Real galaxies have a large range in both mass and size, so that the two model
series discussed above (which assume fixed $M$ and $R$) are highly idealized.
It is inconceivable that two galaxies identical in their mass and size have largely
different collapse time $\beta$ or mass fraction $\Gamma$.
Likely dependence of $\beta$ and $\Gamma$ on the galaxy property should be taken into
account
in order to construct more realistic models.

\subsection{Collapse Timescale: $\beta$}

Infall of the gas from outside the disk plane is a natural consequence of galaxy formation
from extended halos. It has been also introduced into chemical evolution models
to reproduce the metallicity distribution in the solar neighborhood and the age-metallicity
relation (e.g., \cite{lac85}).

It is very difficult to deduce quantitatively the time scale of the protogalaxy collapse
from observational data. However, several correlations observed among spiral galaxy properties,
especially colors and gas contents, provide a
circumstantial
evidence that this timescale, $\beta$, varies from galaxy to galaxy in a systematical way.
 Late-type
disk galaxies 
have a larger mass fraction of the interstellar medium relative
to the galaxy total mass (e.g., \cite{you90,cas98}) and 
 bluer total colors (e.g., \cite{dev74,gav93})
than early-type spirals.
These characteristics indicate that the stellar population in late-type 
galaxies is relatively younger, suggesting a slower build-up of their disks,
in view of the results for the model series A discussed above. 
Although variation along the Hubble sequence appears substantial, dependence on the galaxy
luminosity seems to be much larger.
The well known color-magnitude relation
(e.g., \cite{tul82}) states that the galaxy becomes bluer as it becomes fainter.
Recent analysis by Gavazzi(1993) has found that about two-thirds of the total variation in
spiral galaxy color are due to luminosity difference, and only one-third is contributed 
by the dependence on the Hubble morphological type. The galaxy luminosity has a large
influence also on the gas content.
Gavazzi's (1993) plots for separate morphological classes show that 
at a fixed morphological type the relative gas content
increases by $\sim 10$ as the
galaxy luminosity decreases by 4-5 magnitudes.
Actually, the dependence on the luminosity seems to be stronger than the morphological type dependence, 
being consistent with large scatter of the gas mass fraction at a fixed morphological type seen in Young (1990).

Systematic change of the collapse timescale along the Hubble type
and the galaxy luminosity is plausible also from a theoretical point of view.
A spiral galaxy of an earlier morphology tends to have a higher density of matter
within its optical extent at the same luminosity as suggested by its larger rotational
velocity (e.g., Rubin et al. 1985). It is likely that the collapse timescale is 
significantly governed by the radiative cooling of the primordial gas in the protogalaxy,
and if this is the case, a higher density should have led to a more rapid collapse owing to more
efficient cooling. On the other hand, a smaller galaxy may have been more strongly affected
by the energy injection from internal star formation process than a more massive
galaxy. Then the collapse of a small galaxy must have been prolonged considerably 
by the feedback from the initial star formation [Such feedback mechanisms would have
caused mass loss from the system or even disintegration of the system
in the case of sufficiently intense
star formation as expected in dwarf ellipticals, see Yoshii and Arimoto (1987), for
example]. 

These considerations have led to a specification of the collapse timescale as a decreasing
function of both the galaxy mass and the internal density. The present model assumes
a generalized parameterization of this dependence as follows.

$$ \beta = c ({{M} \over {10^{11}M_{\odot}}})^a ({{\rho} \over {0.1 M_{\odot} pc^{-3}}})^b (Gyr).   \eqno(7) $$

\noindent{Here, the density of the galaxy is defined by
$\rho = M/R^3$, and $c$ is the coefficient to determine the absolute 
value of the collapse timescale. The power indices, $a$ and $b$, determine the steepness
of the dependence, and a larger value adopted for each results in a larger
range in the collapse time.
Both indices are likely to be negative from the discussion above.

 A simple consideration suggests that the range in $\beta$
should be smaller than a factor of ten for the whole population
of disk galaxies. First of all, a collapse time smaller than a few
times $10^8 yr$ will lead to formation of an elliptical galaxy rather than a spiral, 
because such a rapid collapse within a few dynamical times of the system will 
initiate fast star formation which consumes most of the gas before the system
reaches a centrifugal equilibrium.
A plausible upper limit to the collapse timescale comes from the fact that most disk galaxies
appear to have already finished collapse to the disk plane by the present epoch. If the collapse is
still continuing, we should see considerable amount of gas at a large distance from the
galactic plane, which should be emitting X-ray radiation corresponding to the virial 
temperature of the galaxy. $ROSAT$ observations of neaby spiral galaxies (\cite{rea97})
indicate that the amount of such hot gas outside
the galactic plane is less than $\sim 10^9 M_{\odot}$, which is less than 10 percent of 
the galaxy total mass for S0-Sc galaxies in their sample.
This observation, though yet to be extended to a larger sample, seems to preclude a collapse time
which is a large fraction of the age of the universe.

To summarize, I considered two cases as follows.

\noindent{(1) $a=-{{1} \over {2}}$, $b$=0, and $c$=2.0}

\noindent{(2) $a=-{{1} \over {3}}$, $b=-{{1} \over {3}}$, and $c$=2.5}

A steeper power of $a$ and a slightly smaller value of $c$ in case (1) have been taken
to render the range of $\beta$ in the considered ($\rho$, $M$) domain nearly the same 
in both cases.
As an additional constraint, an  upper limit
of 5 Gyr is imposed upon $\beta$.
Actually, the value of $\beta$ is likely to depend also on 
the position in the galaxy (e.g., \cite{mat89,lac85,lar76}).
The value which is used in the present multi-component modeling should be regarded as 
an averaged 
value characteristic of the whole galaxy.

\subsection{Primordial Gas Fraction: $\Gamma$}

Another important parameter, $\Gamma$, i.e., the fraction of mass which eventually
accretes to the galactic plane, is not well constrained from observations, either.
It seems to be  reasonable to assume that $\Gamma$ is equal to the combined mass
fraction (relative to the total galaxy mass )
of the bulge and the disk including interstellar medium,
because there is no firm evidence that at the present epoch
a significant amount of matter resides in the 
halo component except dark matter as stated above. Under this assumption, reliable determination of
$\Gamma$ is still hampered. One ambiguity arises from our poor knowledge about
the mass-to-luminosity ratio for the luminous components. This ratio depends strongly on the
formation history (i.e., time variation of the star formation rate)
and the initiall mass function of the stellar population considered, which are
generally difficult to deduce. 
On the other hand, the total mass, $M$, seems to be well determined from the rotation curve
at least for non-barred galaxies, for which the circular rotation provides a fairly exact
description of the disk kinematics.

Comparing the empirical relation between the mass-to-luminosity
ratio, $M/L_B$, and the $B-V$ color of spirals with the theoretical one predicted by Larson
\& Tinsley's (1978) photometric evolution model, Tinsley (1981) found that the observed increase of
$M/L_B$ with the $B-V$ is much shallower than that theoretically predicted, suggesting that
late-type spirals have relatively more dark matter than early-type ones. Thus, Tinsley (1981)
has claimed that the halo mass fraction is the dominant parameter controlling the
morphological type. Athanassoula et al. (1987) also find that the dominance of the dark
matter is larger in bluer galaxies. However, Jablonka \& Arimoto (1992)
have recently concluded that the halo mass ratio is universal among spiral galaxies
(at least from Sa to Sc), based on a detailed population synthesis analysis treating
the bulge and the disk separately. Also, decomposition models for a selected sample of
spirals having well measured rotation curves seem to show no systematic variation
of this ratio along the Hubble sequence (e.g., \cite{van82,bah85}). 

Persic \& Salucci (1988), on the other hand, conclude that the dark-to-luminous mass ratio
within the optical radius increases as the galaxy becomes less luminous.
According to the result by Ashman (1990) which is based on 
Persic \& Salucci (1988), a galaxy with a mass of $10^{10} M_{\odot}$, for example,
has a twice larger mass fraction of the dark matter on average compared with
the most massive galaxies, although the dispersion at a fixed mass is considerably large
[see Fig.2 of Ashman (1990)].

In view of these arguments, I considered three cases as follows.

\noindent{(1) $\Gamma=0.12 logM-1.00$}

\noindent{(2) $\Gamma=0.15 logM + 0.15 log \rho -2.56$}

\noindent{(3) $\Gamma=0.50$}

\subsection{Model Families}

I have calculated two families of models, differing in the star formation process.
In one family, star formation is allowed to take place always depending on the gas density by taking
$\sigma_{min}=0$. These models are called continuous models. 
A finite threshold, $\sigma_{min}=3(kms^{-1})$,
was specified in another group (hereafter, threshold models).
Each calculated model is specified by three parameters as (c or t)-($\beta$ type)-($\Gamma$ type),
where "c" and "t" denotes continuous and threshold specification for star formation, respectively,
and types for $\beta$ and $\Gamma$ denote the corresponding specifications given 
in \S4.5 and \S4.6, respectively. For example, t-1-3 means the model in which $\beta$ is
 a function only of the mass, $\Gamma$ is constant, and the star formation threshold is introduced.
All the possible combinations of the three parameters have been computed, leading to
12 models in total.
In each model, the galaxy evolution has been calculated at ( $8 \times 8$) grid points on the ($\rho$, $M$) plane,
which are equally spaced both in $log \rho$ and in $log M$. 
The ranges of mass and density are $10^{10}M_{\odot} \leq M \leq
3 \times 10^{12} M_{\odot}$, and $6 \times 10^{-3} M_{\odot}pc^{-3} 
\leq \rho \leq 1  M_{\odot}pc^{-3}$, respectively.
The star formation coefficient was fixed to
$\alpha = 355 M_{\odot} yr^{-1}$ .

In the followings, the results of these models  are confronted with the
currently available observational data.

\section{Observational Material}

Before detailed examination of the model results, this section summarizes the observational data
with which the models are compared.
I confine comparison to the "classical" spirals, i.e., to the morphological type
from Sa to Sc. The observational data, especially those for the bulge-to-disk luminosity
ratio, are well accumulated only for this range of morphology.

\subsection{Bulge-to-Disk Ratio}

Several studies have tried to measure the luminosity ratio of the bulge and disk components
by the decomposition technique (e.g., \cite{yos75,ken85,sim86}).
Figure 8 shows the observed bulge-to-total luminosity ratio as a function of
galaxy mass and density, for the sample used by Whitmore (1984), which is itself based on
the observations by Rubin, Ford, and Thonnard (1980) and Rubin et al. (1982).
I calculated the mass and density for the sample galaxies from the optical radius
(i.e., the radius at which the surface brightness in $B$-band is 25 mag arcsec$^{-2}$),
$R_{25}$, and the rotational velocity, $V_{25}$, at the optical radius as

$$ M = G R_{25} V_{25}^2 $$

\noindent{and}

$$ \rho = M R_{25}^{-3}, $$

\noindent{where $G$ is the gravitational constant.}
Both $R_{25}$ and $V_{25}$ have been taken from Rubin et al. (1980, 1982).
The mass thus calculated will give an approximately correct value for the optical part
of the galaxy, but the density calculated here should be regarded as a rough characteristic 
value for that galaxy, because of a steep density gradient inside the galaxy.

It is recognized in Figure 8 that, on the average,
the  B/T increases as the galaxy mass increases at a fixed galaxy
density, and at a fixed galaxy mass it increases with the density.
However, it is difficult to represent the bulge-to-total luminosity ratio as
a function of one single parameter. The principal component analysis carried out by
Whitmore (1984) indicates that  the surface brightness of the 
galaxy is only one physical quantity (among parameters he investigated)
which shows significant correlation with the B/T.
 Indeed, the lines of constant surface mass density plotted in Figure 8
run roughly in parallel with loci of constant B/T, 
although the irregularity in the distribution of the B/T is large.
In order to see the observed tendency more clearly, Figure 9  divides 
the whole sample into two subgroups according to mass or density, and plots the B/T against
the galaxy mass or the density, separately for each subgroup.
The significance level, $P$, is given, at which the null hypothesis of zero correlation
is disproved as well as the correlation coefficient, $r$.
The correlation is confirmed with a confidence level of $>95$ percent in all cases.
Another compilation by Dale et al. (1997) also supports the dependence of the B/T
on the galaxy mass and the galaxy density displayed in Figure 8.
They do not give the B/T but only the Hubble type for the sample galaxies. 
Figure 10 shows the Hubble morphological type distribution on the ($\rho, M$)-plane
for the Dale et al. (1997) sample.
It is clear that a more massive or denser galaxy tends to have an earlier morphological
type, and that
the regions occupied by galaxies with the same Hubble morphological type 
are also elongated in similar directions to the constant surface density lines.

The observed values of the B/T should be taken with caution. 
The measurement of the bulge-to-disk luminosity ratio
is a very delicate task. Usually a light distribution model composed of
a de Vaucouleurs' bulge (i.e., the $r^{1/4}$-law) and an exponential disk
is fitted to the observed surface brightness profile. However, this method
is criticized recently (e.g., \cite{and94,dej96}) on the ground that
the de Vaucouleurs' density distribution does not decrease sufficiently 
fast as the radius so that the fitting is significantly influenced by the 
irregular light distribution sometimes observed at larger radii.
Andredakis \& Sanders (1994)  propose exponential form for both the bulge and the disk
for better decomposition. The result by de Jong (1996) using this method shows 
a clear trend that the B/T decreases as the morphological type becomes later both
in $B$ and $K$ bands, as expected.
The scatter around the mean relation is large, however. Also the measured B/T
with an exponential bulge is systematically smaller than the one obtained by
the de Vaucouleurs' model by a factor of 3-5. 
It should also be noted that any decomposition technique works best when the 
contributions from the bulge and the disk 
are comparable.
On the other hand, detection of a faint disk in the presence of a luminous bulge or a small bulge embedded in
a bright disk is difficult.

\subsection{Gas Content and Star Formation Rate}

Although the primary aim of the present study is to understand the formation of galactic bulges, any successful model 
of galaxy evolution
should be able to explain the observed trend in the gas content and the star formation activity
among galaxies of different type and luminosity.

Young (1990) shows that the mass ratio of the gas including both neutral and
molecular hydrogens relative to the dynamical mass calculated from the rotation curve
decreases systematically as the morphological type becomes earlier. 
The gas fraction given by Young (1990) ranges from 0.03 for Sa-Sab galaxies to 0.3
for Scd galaxies.
The scatter within the same morphological type is as large as $\sim 10$, however.
The ratio of molecular to neutral hydrogens decreases as the Hubble type becomes later.
Gavazzi (1993) provides extensive data for HI content in disk galaxies. His data show that  
HI flux per unit $H$ flux increases as the morphological
type becomes later as expected. At a fixed type,
HI/$H$ is an increasing function of the $H$-band luminosity of the galaxy,
indicating that less massive galaxies are more gas-rich on the average. 

I take here two samples by Young et al. (1989) and Sage (1993), which are given in a convenient 
form for plotting on the ($\rho$, $M$) plane. Figure 11 plots the mass fraction of the gas
by Young et al. (1989). From their original sample of 182 galaxies, I deleted those galaxies 
which lack mass estimate for either HI or H$_2$. Those galaxies suspected to be in ongoing 
interactions/mergers or in abnormal star formation activity (e.g., Markarian galaxies)
 have also been deleted.
For the remainder, the ratio of the total gas mass to the galaxy mass within the optical
radius has been calculated and plotted on the ($\rho$, $M$) plane.
A global trend is recognized in this figure that the gas mass ratio increases as the galaxy mass increases or the 
galaxy density decreases, though the scatter in the ratio is considerably large.
This systematic variation is even clearer for the sample by Sage (1993), which is
plotted in Figure 12.
The gas fraction is seen to increase monotonically toward the lower-left corner
of the diagram.
Quantitative evaluation of the gas mass fraction needs caution, especially for molecular hydrogens.
The conversion factor from the observed CO luminosity to H$_2$ mass is not known accurately,
and may be different from galaxy to galaxy depending on the metallicity.
Another ambiguity comes from the fact that the neutral hydrogen generally extends much farther
than the optical radius. The procedure taken here tends to overestimate the gas mass ratio 
within the optical radius.

The largest set of star formation rate measurements is given by Kennicutt (1983), who
calculated the SFR for a number of spiral galaxies from $H_{\alpha}$ luminosity.
Figure 13 plots the SFR measured by him on the ($\rho$, $M$) plane.
Only an upper limit of the SFR is given for most Sa galaxies.
The distribution pattern of the SFR is not easy to grasp.

However, the plot of the normalized star formation rate, i.e., SFR/M, on this plane reveals a clear trend
(Fig.14).
The galaxies with higher $SFR/M$ values occupy a lower region in the ($\rho$, $M$)
plane. 
This is qualitatively 
consistent with
the known luminosity effect that the gas fraction increases and the color 
becomes bluer as the galaxy becomes
fainter.
However, there appears to be no clear trend from Kennicutt's (1983) data that
the normalized SFR increases for less dense (hence correspondingly later on average) galaxies.
This seems to be at odds with  
the behavior of the gas fraction
plotted in Figures 11 and 12, which suggests a stronger dependence of the gas content on the
galaxy density
than on the galaxy mass.
This may be due partly to heavy obscuration
of $H_{\alpha}$ emission by dust, which should be more severe in late-type spirals.
These somewhat puzzling results indicate a necessity for more improved and extensive observational
data.

\subsection{Epoch of Bulge Formation}

When and how rapid the galactic bulges have been formed is one of the most important
but unsolved problems regarding disk galaxies.
For example, the age of the Milky Way bulge and the age spread in the 
bulge stars have not yet been tied down 
to sufficient accuracy (e.g., \cite{mat90,hol93,ric96,nor96}).
Although quantitative assessment is difficult, it is possible that the age of the bulge is
different from galaxy to galaxy. Metallicity observations by Jablonka et al. (1996)
show that the value [Mg/Fe] decreases as the bulge becomes fainter. This may indicate that
a smaller bulge possessed in general by galaxies of later morphological
types (see Fig.1 of Jablonka et al. 1996) has been formed in a more extended period, though a
definite answer should wait accurate models of chemical evolution. 

Peletier and Balcells (1996) have recently found that
the optical and near-infrared colors of the bulge  are very similar to those
of the disk 
in a number of spiral galaxies of type S0 to Sbc; A blue bulge is likely to be 
associated with a blue disk. One possible interpretation is that the age of the bulge is
correlated with that of the disk and the age difference is a small
portion of the age of the universe.
At the same time, their data seem to suggest a large
difference in the bulge age among different galaxies.
Fitting of single age, single metallicity stellar population models by Vazdekis et al. (1996)
to the $U-R$ and $R-K$ colors of the observed bulges suggests that a number of bulges can have
ages as young as 4 Gyr and a few bulges may have even younger ages of $\sim 1$ Gyr.
Nevertheless, reliable age determination for old stellar populations is difficult because of
degeneracy of their colors.

Recent imaging observations of the nearby 
galaxy bulges carried out by Carollo et al. (1997) with the HST/WFPC2 provide intriguing results. 
They found that a considerable 
percentage of the observed bulges exhibit irregular morphology, being composed of
a few discrete clumps. Carollo et al. (1997)
 infer that the bulge formation is not a completed process
associated with the initial protogalactic collapse, but may be still ongoing or has been finished only
recently in some galaxies. They propose  a disk origin for the bulge matter, and invoke 
bar structures as a tool driving an inflow of disk matter to the galactic center. 
The observation by Carollo et al. (1997) suggests that the irregular bulges are more frequent in 
spiral galaxies of later morphological types (see their Fig.3), and this may indicate later and/or
more prolonged formation of the bulge in a galaxy of a later type.

\section{Model Results}

This  section describes the model results and attempts comparison with the observational data.
The variation of the collapse timescale, $\beta$, and the primordial gas fraction, $\Gamma$, 
is shown in Figure 15.
Comparison is carried out by using the ($\rho$, $M$) plane. It is permitted in 
fitting the models to the observations to
slide the $\rho$-axis and/or the $M$-axis by a small amount. Because the present analytical model
treats only {\it characteristic} values of the galaxy, there is some ambiguity about galaxy
mass and density. Especially, the density is a poorly defined quantity because of its 
steep variation with the galactocentric radius. Furthermore, even if the galaxy density is 
defined to be an average density within the optical radius (i.e., $R_{25}$ usually),
a discrepancy of this radius between different authors (because of different Hubble 
constants adopted, for example) would be much exaggerated in evaluation of the density, through inverse
cubic dependence of the density on the radius.
These ambiguities should be kept in mind whenever the models are compared with the observations.

\subsection{Bulge-to-Disk Ratio}

All of the twelve calculated models exhibit a variation of the bulge-to-disk ratio as
a function of galaxy mass and density, which is qualitatively consistent with the observation.
Figure 16 plots the mass ratio of the bulge relative to the total luminous matter,
$B/T \equiv m_b/(m_b+m_s)$, for all the models calculated. 
Comparison of these values with
the observed B/T luminosity ratios should be done carefully.
The difference in the mass-to-luminosity ratio in bulges and disks will
not totally justify equating the mass ratio to the luminosity ratio.
Also the observed luminosity ratio depends on the decomposition technique used
(\S5.1). Thus I pay attention primarily to qualitative behavior of the B/T ratio.
The calculated $B/T$ increases as the density and/or the mass of the galaxy increases, showing
the qualitatively same behavior as the observation suggests.
It is encouraging that the
range in the $B/T$ plotted in Figure 16 is roughly equal to the observed range 
given by de Jong's (1996) $K$-band photometry
for galaxies from Sa to Sc.
Although the slopes of B/T contours depend on the adopted specification of $\beta$ and $\Gamma$,
the comparison with Figure 8 does not favor any particular model nor reject it.

One intriguing  effect of the star formation threshold, common to all the threshold models
calculated, is an increase of the bulge-to-disk ratio for least massive galaxies, for which
all the continuous models give smallest B/T ratios. This curious behavior is caused by 
substantial amount of the interstellar gas unused by star formation in these low density galaxies.
This phenomenon may be important in understanding of evolutionary status of extremely
late galaxies, and will be discussed in the forthcoming paper.

Figure 17 plots the $B/T$ against $\beta$. Correlation between the two quantities is not always
tight. Modulation by the galaxy density or the mass fraction, $\Gamma$, is not
neglected in general. Therefore,
the conclusion
by Noguchi (1998) that the bulge-to-disk ratio is determined primarily by the collapse
timescale (i.e., the disk formation timescale) is not supported.
The previous conclusion was derived from more limited calculations in which both the galaxy mass
and the galaxy density are fixed and only the collapse timescale was varied
(such as Series A in \S4.4), and may be errorneous.
Nevertheless, it is true that the accretion timescale has  a strong influence on the resulting
bulge-to-disk ratio.

\subsection{Present Gas Content and Star Formation Rate}

Figure 18 displays the mass fraction of the gas component, $m_g/M$, at the present epoch.
In continuous models,
the collapse timescale $\beta$ governs the gas consumption rate strongly, so that
the present gas fraction is tightly correlated with $\beta$.
Thus the continuous models with $\beta$ dependung only on the galaxy mass
(c-1-1, c-1-2, and c-1-3) exhibit nearly horizontal contours
of $m_g/M$ in the ($\rho$, $M$) plane. This behavior may contradict with the observations by Young et al. (1989)
and Sage (1993), which suggest much steeper contours (see Figures 11 and 12).
On the other hand, the continuous models with $\beta$ being a decreasing function of both
mass and density (c-2-1, c-2-2, and c-2-3)
seem to reproduce the observed trend at least qualitatively. 

One possible problem
with these continuous models is that the absolute value of the gas mass fraction tends to be
considerably smaller than observed, as much as by a factor of
$\sim 10$. This discrepancy may not rule out these models convincingly, however.
There is a substantial discrepancy between the interstellar gas masses determined by
different authors. For example,
Casoli et al. (1998) give the gas fraction of $\sim5 \times 10 ^{-3}$ for Sa galaxies
to $\sim 2 \times 10 ^{-2}$ for Sc galaxies, which is considerably smaller than the values by 
Young(1990) and is in better agreement with the models. 

Introduction of a threshold gas density controlled by a constant minimum velocity dispersion
of the gas disk improves the result remarkably. In this case, the gas content at the present epoch is determined solely
by the galaxy parameters, $M$ and $\rho$ (through the epicyclic frequency), and does not
depend on the star formation history of the galaxy.
This is because the galaxy, in the considered ($\rho$, $M$) domain,
 has already reached to the infall-limited regime of star
formation by the present epoch. The gas mass fraction as a function of $M$ and $\rho$
is the same in all the threshold models, and increases as $M$ and/or $\rho$ decreases, in
qualitative agreement with the observation.
It is seen that inhibition of star formation by the threshold boosts up the present-day 
gas content, and brings the models into a better agreement with the observational data.
The observations by Kennicutt (1989) also suggest that the interstellar gas in many spiral 
galaxies is in the infall-limited regime at present.

The specific star formation rate, $SFR/M$, in the continuous models follows the same behavior
as the gas fraction on the $(\rho,M)$ plane (Fig.19).
Regarding star formation, the continuous models with $\beta$ depending only on the galaxy mass
(c-1-1, c-1-2, and c-1-3)
seem to be in best agreement with the observation by Kennicutt (1983) plotted in Figure 14,
though these models may contradict the observation of the gas content as stated.

The threshold models show sporadic starbursts when the gas density is nearly equal to
the threshold density. This makes the contours of $SFR/M$ 
at the present epoch very irregular on the $(\rho,M)$ plane.
Comparison with the observation is not straightforward. Although the model galaxy exhibits
starbursts, it is premature to conclude that the galaxy really experiences a burst.
A real galaxy will be described as a sum of a number of local regions, over each of which
coherence of evolution is maintained.
The present multi-zone model treats one particular region as the representative
one and traces its time evolution. In other words, this
modelling  does not take into account existence of many local regions
which evolve more or less independently. In real galaxies, random behavior (or phase difference) of 
individual regions will smear out starbursts taking place in local regions, and 
a global starburst will not be realized unless evolution of many regions is 
cynchronized by some mechanism.

\subsection{Formation Epoch of Bulges}

Figure 20 indicates the epoch of bulge formation, $t_{bulge}$.
Here $t_{bulge}$ is defined to be the epoch at which 
half the final bulge mass has been accumulated. The bulge formation epoch in all the calculated
models
becomes earlier as the mass and/or the density of the galaxy increases.
The B/T ratio is anticorrelated with the formation epoch roughly, 
which may agree with the observed decrease
of the [Mg/Fe] ratio as the bulge becomes fainter
(\cite{jab96})
and the preponderance of irregular bulges in late-type spirals (\cite{car97}).
A good positive correlation is found between
$t_{bulge}$ and $\beta$. 
Because $\beta$ also determines the major epoch of disk formation, this correlation may 
explain the similarity in colors between the bulge and the disk in many spiral galaxies
observed by Peletier and Balcells (1996).

The age of the bulge indicated in Figure 19 may be argued to be too large, especially
for less massive and/or less dense galaxies. The present multi-zone
model cannot make allowance for the likely variation of the collapse timescale
at different radii in the galaxy. The value of $\beta$ should be regarded as a kind of
average over the entire disk, and tends to overestimate the actual accretion
timescale in the inner disk, which contributes  to the bulge formation
much. This limitation presumably leads to an overestimate of $t_{bulge}$.
Anyway, measured age difference between the bulge and the disk in a galaxy will depend strongly
on how extended part of the disk is considered, because of a large difference of the accretion
timescale at different radii in the disk component.

\section{Discussion}

It has been demonstrated that the clumpy evolution model can reproduce the observed
variation of the bulge-to-disk ratio among spiral galaxies for a wide range of possibility
which will encompass the real situation.
Nevertheless, the quality of the available observational data and the limitation
in the theoretical modelling seem to hamper further narrowing of the parameter range.
Here I discuss a few implications of the present study and limitations of the models.

\subsection{Appearance of Primeval Disk Galaxies}

Only few studies have explored what primeval galaxies might really look like
(e.g., \cite{mei76,bar87,kat92}).
Theoretical predictions on appearance of young galaxies are becoming increasingly
important, because direct observations of high-redshift galaxies enabled by HST
and other new instruments start to provide powerful constraints on theoretical
evolutionary models from morphological and 
dynamical viewpoints.

The CDM cosmogony predicts clumpy appearance 
of primeval galaxies as a direct consequence of dominant small-scale density perturbations imposed
on the matter distribution in the early universe. Indeed, Baron and White (1987)
demonstrate by numerical simulations that a young {\it elliptical} galaxy should not
be observed as a bright single body but as a conglomeration of several discrete
blobs connected by a common faint envelope.
Katz's(1992) dissipational formation model for a spiral galaxy, including star formation
process, also produces clumpy appearance at high redshift, because of imposed initial density
perturbations. In this case, the clumps form during the collapse of the galaxy.

In contrast to these CDM-based simulations, the clumps advocated in the present study
have  no causal relationship with initial density perturbations in the universe.
The clumpy nature of primeval disk galaxies in the present model originates in
the gravitational instability of gas-rich galactic disks formed in the early phase 
of disk galaxy evolution. It is encouraging that HST and large ground-based telescopes
have recently found clumpy structures in a number of high-redshift galaxies, 
although they may be manifestation of CDM clumps. The number of clumps
contained in one object is a few usually, though this number may be severely affected
by the spatial resolution and the surface brightness limit of the instrument used.
Even chain galaxies (e.g., \cite{cow95}), which are elongated objects 
containing several bright blobs, may be primeval disk galaxies viewed edge-on, 
in which the clumps scattered in the disk component are viewed within the projected disk plane 
[Another possibility is suggested by Dalcanton and 
Shectman (1996) that they represent edge-on low surface brightness  
galaxies].
Head-tail systems called "tadpole" galaxies by van den Bergh et al.(1996) may not be a
rectilinear object but edge-on 
manifestation of a clumpy disk in which one of several clumps is particularly
large.

Another possible evidence for clumpy structures other than from direct imaging
comes from the analysis of
correlation functions. Infante et al. (1996) found, for galaxies with 
the average redshift of $<z> \sim 0.35$,
a discontinuity of the correlation function at the separation  of $\sim 6$ arcsec.
This suggests strong clustering of faint galaxies within $\sim 20$ kpc of individual galaxies,
which may be caused by clumping in a single galactic disk. 
Because the mean redshift of their sample is relatively small, it is desirable to extend similar
analysis into a larger redshift.

One caveat in interpreting images of distant galaxies is that
morphology of objects at large redshift is strongly influenced by $k$-correction and
the steep dependence of surface brightness on redshift (of the form $(1+z)^{-5}$).
Bohlin et al. (1991) and Giavalisco et al. (1996a) have cautioned that
those clumpy structures observed in medium to
high redshift galaxies may not be the features characteristic of
early evolution phases but
simply be an exaggerated manifestation of the irregular distributions of star forming regions
such as observed in some galaxies (usually of  late types) in the local universe.

Interactions or mergers with smaller satellite
galaxies are sometimes invoked in the interpretation of peculiar
appearance of high redshift galaxies (e.g., \cite{gri94,van96}).
The most straightforward and powerful test to discriminate between the instability hypothesis 
proposed here and 
the merger/interaction scenario  
is to examine the kinematics of "satellites". In the merger/interaction scenario, we expect 
random orientation of clump orbits  relative to the primary because there is no reason to
consider that the bombardment of other galaxies from outside has any
preferred geometry with respect to the primary.
On the other hand, 
it is inevitable in the present scenario
that the motions of "satellites" are co-planar and all the satellites rotate 
in the same direction around a common center.
Spectroscopic observations will provide a direct check.

\subsection{Creation of the Bulge-Disk Structure}

The traditional Hubble classification scheme is based on three 
morphological characteristics; 
prominence of the bulge relative to the disk, the pitch angle of the spiral arms, 
and the degree of resolution of the arms into stars ( Sandage 1961).
 However, Sandage et al. (1970) argue that the most fundamental
parameter distinguishing between different Hubble types is the relative gas content,
presumed to have been determined at the time of formation. 
Yoshizawa \& Wakamatsu (1975) found that the relative prominence of the disk and
bulge components is well correlated with the morphological type, and argue that
disk galaxies are specified quantitatively by both
the bulge-to-disk ratio (as a morphological indicator) and the luminosity of the disk
(as a scale indicator).
Van den Bergh (1976) has proposed a two-dimensional classification scheme in which the 
bulge-to-disk ratio and the gas content are independent parameters, with the former determined
by the formation process while the latter changeable due to environmental factors.
These propositions are, however, made on more-or-less phenomenological grounds and do not always
address in what manner these fundamental quantities (such as the gas content and the bulge-to-disk
ratio) have been determined in the evolution of a particular galaxy. 
More systematic and quantitative attempts include application of the principal component 
analysis to determine the number of independent parameters defining
galaxies. Brosche (1973) and other groups
have concluded that just two dimensions could explain most of the 
observed diversity in spiral galaxies.

Dynamical study of the formation and evolution process of galaxies has a long
history (e.g., \cite{egg62}).
One of the most important goals in this area is to understand origins of the observed
variety of galaxies.
A series of numerical works carried out by Larson (1969, 1974, 1975, 1976)
stand out as a landmark in the theoretical galactic astronomy. In his models,
the formation of a disk galaxy proceeds in two stages characterized by star formation
processes that operate at very different rates. First, there is a rapid star formation process that forms a spheroidal 
component, and later a much slower star formation
permits most of the residual gas to condense into a disk before it is consumed by star formation.
Larson (1976) envisages two possibilities as the cause of the reduction of the star
formation efficiency after the bulge formed, which is required for the formation of a
significant disk component. The first is inhibition of star formation by the tidal force 
exerted on the gas clouds by the already existing bulge. Second possibility is
that the protogalactic gas has a two-phase structure with dense clouds that rapidly
form stars in a spheroidal component and less dense intercloud gas which does not form
stars until it has settled to a disk.
He has also prescribed turbulent viscosity which is large in the early phase of evolution when the 
bulge grows but is reduced in later phases, 
in order to get distinct separation between the disk and the bulge.
Using these settings, Larson (1976) has reached to the conclusion that the bulge-to-disk
ratio depends mostly on the star formation rate, which is in turn controlled by the initial
density or velocity dispersions in protogalaxies.
	
In marked contrast with Larson's (1976) model, the present study proposes a more disordered
and chaotic
formation of disk galaxies. The dominance of massive subgalactic clumps and resulting
dynamical processes  constitute the main
ingredient of the present model. The bulge is assembled from those clumps formed
by the local gravitational instability in the disk component.
The model appears not to require two star formation processes
operating at different timescales, unlike
the models constructed by Larson (1976), though more realistic simulations would be
necessary to establish this point.
In the present model, the clumpy nature of the young gas-rich disk provides the required
viscosity to form a bulge (e.g., \cite{lin87}). Larson (1976) had to introduce viscosity
{\it a priori} because the axisymmetry of the configuration imposed on his models could not 
allow radial transport of angular momentum by non-axisymmetric perturbations
such as clumps, bars, and spiral arms.
Although the present study stresses the secondary nature of the galactic bulges, this does not
mean that the {\it whole} disk has formed before the bulge, as repeatedly 
cautioned. Only inner parts of the disk
contribute to bulge formation, and outer parts are considered to form much later than
the bulge by a slow accretion of the primordial gas from the outer halo.
 In this respect, the present scenario is not  so drastically  different from Larson's (1976) model
as it might seem.

\subsection{Longevity of Clumps}

The present study stresses the importance of heavy clumps formed in the early galactic
disks in driving long-term disk galaxy evolution.
Therefore the longevity of these clumps is a key factor of the present model.
Although the present study takes into account the contribution from the formed clumps
of all the mass scales, it may be considered that clumps with smaller masses are prone to
destruction in general. The possible destruction processes include energy injection
from internal star formation and tidal destruction due to the galaxy gravitational
field. 
Strong concentration of the gas clouds into narrow spiral arms observed in 
nearby galaxies points to relatively short lifetimes for those clouds.
For example, the giant molecular clouds , with masses of $\sim 10^6 M_{\odot}$,
are considered to have a lifetime shorter than a few times $10^8yr$ , perhaps due
to energy deposit from young massive stars born in them. Even giant molecular associations
(GMAs), with estimated masses of several times $10^7 M_{\odot}$, seem to be transient. Rand \&
Kulkarni (1990) found that GMAs in the inter-arm regions of M51 are gravitationally unbound, and
argued that they may be disrupting  due to tidal shearing by the background gravitational
field of the galaxy. If this is correct, the estimated lifetime of these GMAs is several times
$10^8$ yr.
The present model sometimes develops clumps with  masses of $10^{8-9} M_{\odot}$.
Fate of these extremely massive clumps is not clear because they are absent
from nearby (i.e., present-day) galaxies. Such clumps may also suffer 
from disintegration.

Then is the treatment in the present study totally unrealistic?
Answer is probably 'No'.
Assuming that a clump of any mass has a finite lifetime, probably depending on
its internal structure and the strength of the external tidal field, the interstellar gas 
in real galaxies
will be circulated through different phases as follows. First , a group of clumps will be formed
owing to gravitational instability in the gas disk. Untill these clumps are destroyed by some
mechanism, they will feel dynamical frictions and move inward to the galactic center.
When the clumps are destroyed, clump material
is dispersed into the interstellar space and 
the inflow driven by dynamical friction
 stops. However, after a certain period, a new generation
of clumps are formed from this diffuse material by gravitational instability, and they
resume inward motion. Actually, this recycling will not be a coherent process over the entire
galactic disk, but every local region in the disk will experience recycling with its own phase, 
independently of each other.

The numerical simulation described in \S2 and \S3 differs from
the situation considered here despite that it does include energy feedback from
star formation events in a simple form.
The massive clumps formed in the numerical model maintain their identity
untill they merge with other clumps or they are swallowed by the bulge.
On the other hand, the analytical multi-component models presented in
\S4 seem to better fit in with the circulation picture of the interstellar medium.
They calculate the typical mass of clumps from the instantaneous
gas surface density. Therefore, the clump mass is changing as a function of time.
In other words, the models do not assume the identity of each clump for the whole period
of galaxy evolution.
Then, those analytic models provide a fairly good description of
actual situation, provided that the average lifetime of clumps is a significant
fraction of the whoel period of one circulation cycle. 

\subsection{Undiscussed Issues}

The present study has mainly discussed a limited range of the whole family of spiral galaxies,
namely, from type Sa to Sc. These galaxies occupy the upper right half of the ($\rho, M$)
plane employed here. Compared with these galaxies, relatively little is known 
about galaxies later than Sc, which occupy the lower left part of the ($\rho, M$)
diagram. The present study may provide an interesting implication for these galaxies
as stated in \S6.1,
but it will be treated in a separate paper.

Another issue which may be related to the problem of extremely late galaxies
is the evolutionary status of the low surface brightness galaxies.
Recent observations have been providing
 a mounting evidence that the realm of galaxies is actually dominated by very faint
galaxies which have eluded past observations.
The universality of the disk central surface brightness (\cite{fre70}) is turning out to be 
due to selection effect, and those galaxies with a central brightness fainter than 
$\sim 23 B$ mag arcsec$^{-2}$, which are
often called low surface brightness (LSB) galaxies (\cite{imp97} and references therein),
may occupy as much as half the total galaxy population.
LSB galaxies tend to be of a late
morphological type (\cite{sch92}), though early-type analogues are
 sometimes observed (\cite{spr95}).
One possibility is that most LSB galaxies represent a continuation of disk galaxies into
morphological types later than Sd and Sm.
Protogalactic systems with a high angular momentum and/or a low mass are suggested to 
be precursors of LSB galaxies (Dalcanton et al. 1997). 
 Global characteristics of LSB galaxies have been investigated
intensively in recent years (e.g., \cite{deb95,deb96,deb97}). Most LSB galaxies appear to lack
any significant bulge component. However, interestingly enough,
Sprayberry et al. (1995) report several giant LSB
galaxies possessing a large bulge component.

 It is not clear at present how these LSB galaxies
can be incorporated into the galaxy evolution model proposed here. Since the present model 
is a model of bulge formation largely,  more detailed information on the bulge-to-disk
ratio for LSB galaxies is necessary to examine whether the present model can accomodate
the class of LSB galaxies as its part or not.

\section{Conclusions}

Collapse of a protogalaxy composed of dark matter and primordial gas has been investigated by
numerical simulations and analytical multi-zone modelling in an attempt to examine early
evolution of disk galaxies. Importance of the ample interstellar matter existing in 
young galactic disks has been highlighted.
Confrontation of the theoretical results with the available observational data has led to a
new picture for disk galaxy evolution, in which the bulge is the secondary 
object formed from disk matter.

As the protogalaxy collapses, a gaseous disk starts to form. 
Very few stars form before the disk formation because the gas density is low. The formation
of the disk causes drastic increase in the gas density and initiates star 
formation process. The galactic disk in early evolution phases is rich in the interstellar gas
and the efficient energy dissipation keeps the disk dynamically cold.
Then the gravitational instability
sets in,
leading to the formation of massive clumps rotating in the disk plane. 
The individual mass of these clumps can be as large as  $\sim 10^9 M_{\odot}$.
Intense star formation occurs in these clumps.
Therefore, the galaxy at this epoch exhibits a clumpy and irregular appearance in optical
wavelengths,
which may give explanation of morphologically peculiar galaxies observed at high redshift.
While orbiting in the disk plane, the clumps tend to merge with each other and
make successively larger clumps.
The clumps loose their orbital kinetic energy through dynamical friction against surrounding stars
and gas clouds, and accumulate to the central region, thus forming a spheroidal bulge.
The collapse timescale of the protogalaxy will be larger in its outer parts.
Slower accretion of the primordial gas is considered to have established outer parts of the disk
after the bulge formation is mostly finished.

Simple analytical models have been constructed, in which a disk galaxy is described as a multi-component
system comprising a dark halo, gaseous and stellar disks, and a bulge. Evolution of these components
is controlled by two parameters; the accretion timescale,
i.e., the rapidness with which the primordial gas contained in the halo region
accretes to the disk plane, and the mass fraction of the primordial gas at the beginning.
Based on the observational evidence,
the accretion timescale in real spiral galaxies has been assumed to be a decreasing function
of both the galaxy mass and the galaxy density, whereas the fraction of 
the primordial gas has been assumed to 
increase as the galaxy mass and/or the galaxy density increases.
Under this specification, the analytical modelling
 of the clumpy galaxy evolution picture
has succeeded in reproducing the observational result that
a galaxy having a large total mass and/or a large internal density tends to have a large
bulge-to-disk ratio,
which is established at least for the morphology range 
of Sa to Sc.
This success suggests importance of the clumpy evolution in young disk galaxies,
though the evolution of real spiral galaxies may be a more complicated process comprising
several fundamentally different mechanisms.

\clearpage

\appendix

\begin{center}
NUMERICAL EVALUATION OF SINKING TIMESCALE OF A CLUMP
\end{center}

A massive body orbiting in the disk component of a galaxy spirals in to
the galactic center by the action of dynamical friction.
The timescale of this inward motion of the clump has been evaluated by numerical simulations as follows.
Parameters of each simulation are listed in Table 1.

N-body models for a disk galaxy, which consists of a halo and a disk, 
have been constructed first. The models contain no interstellar gas.
Here, the halo is not meant to represent the entire (dark) halo
which might surround the visible galaxy, but the portion of the 
dark halo {\it inside} the optical radius plus any luminous
spheroidal components such as a bulge.
The total mass of the galaxy is unity and the mass of the disk component is $m_d$.
Both components are truncated at the galactocentric radius of unity.
The halo and the disk are constructed by 5000 and 50000 collisionless particles,
respectively. The gravitational softening radius is 0.04 and
0.02 for the halo and disk particles, respectively.

The model is fundamentally based on Fall \& Efstathiou (1980).
The stellar disk has an exponential surface density distribution
with a scale length of 0.25, in agreement with observations.
The disk rotates nearly rigidly in the inner parts and at a nearly constant
velocity in the outer parts. The turn-over radius which divides these two parts
is denoted by $r_m$. Thus the global shape of the rotation curve is determined by specifying $r_m$.
The halo is assumed to be spherically symmetric, and its volume density distribution
is determined so that the rotational velocity in the disk plane due to the combined 
gravitational field of the halo and the disk matches the specified rotation curve.

The velocity dispersion of the halo is chosen as follows.
First, the isotropic velocity dispersion at each radius is
calculated so that the condition
for "local virial equilibrium" is satisfied everywhere
(see Noguchi 1991 for details). A trial simulation showed that such a condition
does not lead to virial equilibrium for the entire system. 
In order to alleviate this, the velocity dispersion
was multiplied by a factor of 0.65.

This disk galaxy model is evolved in isolation before the sinking simulations.
First only the halo component is evolved for 15 dynamical times
with the disk component fixed.
After the halo is relaxed, the disk is activated. At this time, the gravitational 
force acting on each disk particle is calculated and the circular velocity is
given to that particle so that the centrifugal force is exactly balanced
with the gravity. Next, small random velocities are given to
disk particles , which correspond to a specified $Q$ parameter of 
Toomre (1964). The rotational velocity of each star is then corrected for the 
contribution from this random motion.
This state just after the activation of the disk is adopted as the initial
condition for the disk galaxy in the sinking simulations.

The clump is treated as a particle which has a mass, $m_{cl}$, and a gravitational 
softening radius of $r_s$, which is regarded as the effective radius
of the clump. 
The mass and radius of the clump given in Table 1 are also in units
of those of the disk galaxy.
The clump starts at the galactocentric radius $r=0.8$ in the disk plane.
The initial velocity is that of the circular motion, and the orbital motion is in the same direction
as the disk rotation. This initial condition is appropriate for the clumps formed from the 
disk material. A number of simulations have been carried out by varying
$m_d$, $m_{cl}$, $r_m$, and $Q$.
Each simulation is performed untill the galactocentric radius of the clump decreases to 
0.2. Epoch of this moment is taken to be the dynamical friction timescale, $\tau_{fri}$,
and listed in Table 1.
Gravitational interactions between all the particles have been calculated by the tree code
(e.g., \cite{bar86}), and the orbit integration has been done by the leap-frog scheme with a
time step of 0.01, i.e., one hundredth of the dynamical timescale (A time step of 0.015 was used in
the halo relaxation phase).

\clearpage

\begin{deluxetable}{ c  c  c  c  c  c  c }
\footnotesize

\tablecaption{ Timescale of dynamical-friction induced sinking.\label{tbl: param_of_model}}


\tablehead{
\colhead{Model} & \colhead{\(m_{d}\)} & 
\colhead{\(m_{cl}\)} & \colhead{\(r_{m}\)} & 
\colhead{\(Q\)} & \colhead{\(\tau_{fri}\)\tablenotemark{a}}  &
\colhead{note\tablenotemark{b}}
}

\startdata
1 & 0.3  & 0.01  & 0.3 & 1.5 & 5.41 & \nl
2 & 0.1  & 0.01  & 0.3 & 1.5 & 11.4 & \nl
3 & 0.2  & 0.01  & 0.3 & 1.5 & 7.32 & \nl
4 & 0.3  & 0.03  & 0.3 & 1.5 & 3.78 & \nl
5 & 0.3  & 0.003  & 0.3 & 1.5 & 11.1 & \nl
6 & 0.3  & 0.001  & 0.3 & 1.5 & 21.7 & \(r_{s}\)=0.02 \nl
7 & 0.3  & 0.01  & 0.1 & 1.5 & 5.31 & \nl
8 & 0.3  & 0.01  & 0.7 & 1.5 & 6.32 & \nl
9 & 0.3  & 0.01  & 0.3 & 1.0 & 5.29 & \nl
10 & 0.3  & 0.01  & 0.3 & 2.0 & 5.96 & \nl
11 & 0.3  & 0.01  & 0.3 & 1.5 & 4.75 & \(r_{s}\)=0.02 \nl
12 & 0.3  & 0.01  & 0.3 & 1.5 & 4.92 & \(r_{s}\)=0.04 \nl
\enddata

\tablenotetext{a}{The timescale, $\tau_{fri}$, is the time it takes the clump to move
from $r=0.8$ to $r=0.2$, and its unit is the dynamical time of the galaxy.}

\tablenotetext{b}{The effective radius of the clump, $r_s$, is 0.07 unless specified.}

\end{deluxetable}

\figcaption[f1.eps]{Morphological evolution of the numerical model.
 (a) The projection  of the gas cloud particles onto the
x-y plane (i.e., the disk plane).
Only half the particles selected at random are displayed. 
Time, $t$, in units of Gyr is indicated in the upper right corner 
of each frame. 
Coordinates are given in units of kpc.
(b) The same as (a)
but the projection onto the x-z plane.
The x-y and x-z projections of all the stellar particles younger than 
$10^7$ yr are given in (c) and (d), respectively. The stars older than
$10^7$ yr are displayed in (e) and (f).
In panel (e), the five massive clumps used for subsequent alanyses are indicated
at $t$=0.64 Gyr, enclosed by circles. \label{fig1}}

\figcaption[f2.eps]{The mass fractions of the gaseous and total disks
with respect to the total mass in the numerical model (solid lines).
The fraction of the gaseous disk, $f_g$, is a summation over
the gas particles with $|z-z_c|<0.1 R$ and $0.3 R <r<1.5 R$,
where $z_c$ denotes the z-coordinate of the mean disk plane at each time and $r$
is the cylindrical radius measured from the z-axis.
The mass fraction of the total disk, $f_d$, is the sum of $f_g$ and 
the mass fraction of the stellar  disk.
The latter is summed over all the stellar particles with $0.3 R <r<1.5 R$.
The left panel indicates the evolutionary track on the $f_d-f_g$ diagram, whereas the time evolution
of $f_g$ is shown in the right panel. 
Time, $t$, is given in units of Gyr here.
The dashed lines indicate the evolution of an analytical multi-component model
explained in \S4. The analytical model plotted here has the same 
mass and radius as the numerical model. The fraction of the primordial gas is also the same 
as in the numerical model (i.e., $\Gamma=0.5$), and the collapse timescale was set to $\beta=$ 0.3 Gyr.
The dynamical friction timescale 
given in eq(5) of the text was multiplied by 0.3 in this model.
Open circles in the right panel indicate the expected mass of the individual clumps
(in units of $M_{\odot}$)
in the analytical model, which is evaluated as explained in \S4.3.
 \label{fig2}}

\figcaption[f3.eps]{Time evolution of the star formation rate in the numerical model
(solid line).
 Time, $t$, and the star formation rate
are given in units of Gyr and $M_{\odot}yr^{-1}$, respectively.
The dashed line indicates the SFR in the analytical model plotted in Figure 2.
 \label{fig3}}

\figcaption[f4.eps]{Distribution of the age for the stars contained in the bulge region in 
the numerical model.
The center of the bugle is defined to be that of the most massive clump.
Each histogram shows, for the given epoch, the relative number of the stars
which were born at the time indicated in the abscissa.
Solid lines are distributions for stars located within 0.1 $R$ of the bulge center,
whereas the dotted lines for stars located within 0.025 $R$. 
Time is in units of Gyr.
\label{fig4}}

\figcaption[f5.eps]{Sinking of a heavy clump through the stellar disk of model spiral
galaxies. Details of the numerical calculation are given in Appendix. The ordinate
indicates the galactocentric distance of the clump in units of the radius of the model
galactic disk. The abscissa is the time normalized by the dynamical friction timescale,
$\tau_{fri}$, namely the time which takes the clump to move from its initial radius of 0.8 
to the final radius of 0.2.
Model parameters are given in Table 1., but individual models are not discriminated 
in this plot.
 \label{fig5}}

\figcaption[f6.eps]{Time evolution of the models in Series A.
All the models have the same mass, $M=10^{11}M_{\odot}$, and the same radius, $R=10$kpc.
Only the accretion timescale, $\beta$, has been varied.
From the most rapidly increasing curve, $\beta=$ 0.5, 1.08, 2.32, and 5.0 Gyr, in all the panels.
All the models have $\Gamma$=0.5.
The dynamical friction timescale 
given in eq(5) of the text was multiplied by 0.3 in all the models.
The star formation rate, SFR, and the mass of the clump, $m_{cl}$, are in
units of $M_{\odot}yr^{-1}$ and $M_{\odot}$, respectively.
 The mass ratio of the bulge to the total luminous matter, B/T,
is defined by $m_b/(m_b+m_s)$.
 \label{fig6}}

\figcaption[f7.eps]{Same as Figure 6 but for the models in Series B, in which
only $\Gamma$ is varied. Four models are indicated here, with
$\Gamma$=0.1, 0.17, 0.29, and 0.5, from the thinnest to progressively thicker
solid curves. All the models have $\beta$=2 Gyr. 
The dynamical friction timescale 
given in eq(5) of the text was multiplied by 0.3 in all the models.
The dashed line indicates, for each model, the quantity divided by $\Gamma$ of that
model. Again, $\Gamma$ increases as the line becomes thicker. For these normalized
quantities, the ordinate is arbitrary, to facilitate only relative comparison
between different models.
Note that the normalization makes time variation in the masses of the stellar and gaseous disks
($m_s$, $m_g$) and the star formation rate (SFR) almost identical in different models, 
indicating that these three quantities are scaled just in proportion to the mass fraction,
$\Gamma$, of the initial primordial gas. 
On the other hand, the clump mass ($m_{cl}$) and the bulge mass ($m_b$) show strongly
nonlinear dependence on $\Gamma$, which leads to largely different final values of B/T.
 \label{fig7}}

\figcaption[f8.eps]{The bulge-to-total luminosity ratios for a sample of
spiral galaxies with morphological type of Sa to Sc, taken from Whitmore (1984).
The ordinate indicates the mass, $M$, of the galaxy inside the optical radius 
in units of $M_{\odot}$, whereas the abscissa indicates the density, $\rho$, within the
optical radius in units of $M_{\odot} pc^{-3}$.
The area of each circle is proportional to the B/T ratio for the
corresponding galaxy.
The dashed lines specify the constancy of the surface density defined
by $\Sigma \equiv M/R^2$, where the mass, $M$, is in units of
$M_{\odot}$, and the optical radius, $R$, is in units of parsec. \label{fig8}}

\figcaption[f9.eps]{Correlation of the bulge-to-total luminosity ratio (B/T)
and the mass or the density for Whitmore (1984)
galaxies plotted in Figure 8. The whole sample is divided into two groups according to the mass
or the density, and correlation is examined between the B/T and the other
quantity in each subsample. 
The correlation coefficient, $r$, and the probability for the null hypothesis
of zero correlation, $P$, are indicated. \label{fig9}}

\figcaption[f10.eps]{Distribution of galaxies with different morphological types on the ($\rho$, $M$)
plane. 
The galaxy sample is taken from Dale et al. (1997).
The ordinate indicates the mass, $M$, of the galaxy inside the optical radius 
in units of $M_{\odot}$, whereas the abscissa indicates the density, $\rho$, within the
optical radius in units of $M_{\odot} pc^{-3}$.
The dashed lines specify the constancy of the surface density defined
by $\Sigma \equiv M/R^2$, where the mass, $M$, is in units of
$M_{\odot}$, and the optical radius, $R$, is in units of parsec. \label{fig10}}

\figcaption[f11.eps]{Mass fraction of the interstellar gas realtive to the total matter
inside the optical radius of the galaxy.
The galaxy sample is taken from Young et al. (1989).
The ordinate indicates the mass, $M$, of the galaxy inside the optical radius 
in units of $M_{\odot}$, whereas the abscissa indicates the density, $\rho$, within the
optical radius in units of $M_{\odot} pc^{-3}$.
Here the interstellar gas means the sum of the molecular gas and the atomic gas.
The area of the circle is proportional to the gas mass fraction.
Dotted circles mean upper limits in either molecular mass or atomic mass. \label{fig11}}

\figcaption[f12.eps]{Same as Figure 11, but for the sample by Sage (1993).
Dotted circles mean upper limits in  molecular mass. \label{fig12}}

\figcaption[f13.eps]{Star formation rate for the sample of spiral galaxies
compiled by Kennicutt (1983). Only galaxies with type Sa to Sc are plotted.
The ordinate indicates the mass, $M$, of the galaxy inside the optical radius 
in units of $M_{\odot}$, whereas the abscissa indicates the density, $\rho$, within the
optical radius in units of $M_{\odot} pc^{-3}$.
The data used to calculate the mass and the density are taken from RC3 (de Vaucouleurs et al.
1991).
Crosses, which correspond to Sa galaxies mostly,
 mean non detection of measurable star formation activity.
Breakdown of the whole sample by the SFR (in units of $M_{\odot}yr^{-1}$)
is given in other five panels. \label{fig13} }

\figcaption[f14.eps]{Same as Figure 13, but for the specific star
formation rate, SFR/M, for the sample of spiral galaxies
compiled by Kennicutt (1983). Only galaxies with type Sa to Sc are plotted.
Here, the specific SFR is defined to be the star formation rate divided by
the galaxy mass, $M$, inside the optical radius, and its unit is yr$^{-1}$.
Crosses, which correspond to Sa galaxies mostly,
 mean non detection of measurable star formation activity. \label{fig14}}

\figcaption[f15.eps]{The accretion timescale, $\beta$, and the mass fraction of the primordial gas,
$\Gamma$, in the analytical multi-zone models.
The model name (see text for explanation) is indicated in the upper right corner
of each panel.
The ordinate indicates the mass, $M$, of the galaxy inside the optical radius 
in units of $M_{\odot}$, whereas the abscissa indicates the density, $\rho$, within the
optical radius in units of $M_{\odot} pc^{-3}$.
The dotted contours correspond to  $\beta$=1, 2, 3, 4, and 5 Gyr,
from the thinnest to progressively thicker lines in all the panels.
The solid contours correspond to $\Gamma$= 0.25, 0.3, 0.35, 0.4, and 0.45,
from the thinnest to progressively thicker lines in all the panels.\label{fig15}}

\figcaption[f16.eps]{Same as Figure 15, but for the ratio of the bulge mass to the 
total luminous mass at the present epoch, defined by $B/T=m_b/(m_b+m_s)$, where $m_b$ and $m_s$ are 
masses of the bulge and the disk, respectively.
The contours are for log(B/T)=-0.5, -0.75, -1.0, -1.25, and -1.5,
from the thickest to progressively thinner lines in all the panels.\label{fig16}}

\figcaption[f17.eps]{The bulge-to-luminous mass ratio, B/T, plotted against the accretion
timescale, $\beta$ (in Gyr), for each model.\label{fig17} }

\figcaption[f18.eps]{Same as Figure 15, but for the ratio of the gas mass to the 
total galaxy mass, $m_g/M$, at the present epoch.
The contours are for $m_g/M$=0.001, 0.002, 0.004, 0.008, 0.016, 0.032, and 0.064,
from the thinnest to progressively thicker lines in all the panels.\label{fig18}}

\figcaption[f19.eps]{Same as Figure 15, but for the ratio of the star formation rate to the 
total galaxy mass, $SFR/M$, at the present epoch.
The contours are for $SFR/M$= (0.5, 1, 2, 4, 8, 16, 32, 64)$\times 10^{-12}yr^{-1}$,
from the thinnest to progressively thicker lines in all the panels.\label{fig19}}

\figcaption[f20.eps]{Same as Figure 15, but for the epoch of bulge formation,
$t_{bulge}$, which is defined to be the epoch where the bulge mass reached to half
of the present value.  
The contours are for $t_{bulge}$= 1, 2, 3, 4, and 6 Gyr,
from the thinnest to progressively thicker lines in all the panels.\label{fig20}}


\clearpage

\begin{thebibliography}{}

\bibitem[Abraham et al.\ 1996]{abr96} Abraham, R.G., Tanvir, N.R., Santiago, B.X., Ellis, R.S., 
Glazebrook, K., and van den Bergh, S. 1996, \mnras, 279, L47

\bibitem[Andredakis and Sanders 1994]{and94} Andredakis, Y.C., and Sanders, R.H. 1994, \mnras, 267, 283

\bibitem[Ashman 1990]{ash90} Ashman, K.M. 1990, \apj, 359, 15

\bibitem[Athanassoula et al.\  1987]{ath87} Athanassoula, E., Bosma, A.,
and Papaioannou, S. 1987, \aap, 179, 23

\bibitem[Bahcall and Casertano 1985]{bah85} Bahcall, J.N., and Casertano, S. 1985, \apj, 293, L7

\bibitem[Barnes and Hut 1986]{bar86}
Barnes,J.E., and Hut,P. 1986, Nature, 324, 446

\bibitem[Baron and White 1987]{bar87} Baron, E., and White, S.D.M. 1987, \apj, 322, 585


\bibitem[Bohlin et al.\ 1991]{boh91} Bohlin, R.C. et al. 1991, \apj, 368, 12


\bibitem[Brosche 1973]{bro73} Brosche, P. 1973, \aap, 23, 259 

\bibitem[Carollo et al.\ 1997]{car97} Carollo, C.M., Stiavelli, M., de Zeeuw, P.T., and Mack,
J. 1997, \aj, 114, 2366

\bibitem[Casoli et al.\ 1998]{cas98} Casoli, F. et al. 1998, \aap, 331, 451


\bibitem[Cole et al.\ 1994]{col94} Cole, S., Aragon-Salamanca, A., Frenck, C.S., Navarro, J.F., and Zepf, S.E.
1994, \mnras, 271, 781


\bibitem[Cowie et al.\ 1995]{cow95} Cowie, L.L., Hu, E.M., and Songaila, A. 1995, \aj, 110, 1576


\bibitem[Dalcanton and Shectman 1996]{dal96} Dalcanton, J.J., and Shectman, S.A. 1996, \apj, 465, L9

\bibitem[Dalcanton et al.\ 1997]{dal97} Dalcanton, J.J., Spergel, D.N., and Summers, F.J. 
1997, \apj, 482, 659




\bibitem[Dale et al.\ 1997]{dale97} Dale, D.A., Giovanelli, R., Haynes, M.P., Scodeggio, M.,
Hardy, E., and Campusano, L.E. 1997, \aj, 114, 455


\bibitem[de Blok et al.\ 1995]{deb95} de Blok, W.J.G., van der Hulst, J.M., and Bothun, G.D.
1995, \mnras, 274,235


\bibitem[de Blok et al.\ 1996]{deb96} de Blok, W.J.G., McGaugh, S.S., and van der Hulst, J.M.
1996, \mnras, 283, 18

\bibitem[de Blok and McGaugh 1997]{deb97} de Blok, W.J.G., and McGaugh, S.S.
1997, \mnras, 290, 533


\bibitem[de Jong 1996]{dej96} de Jong, R.S. 1996, \aap, 313, 45


\bibitem[de Vaucouleurs 1974]{dev74} de Vaucouleurs, G. 1974, in IAU Symp. 58, The Formation
and Dynamics of Galaxies, ed. J.R. Shakeshaft 
(Dordrecht: D.Reidel Publishing Company), 1

\bibitem[de Vaucouleurs et al.\ 1991]{dev91} de Vaucouleurs, G., de Vaucouleurs, A.,
Corwin, H.G., Jr., Buta, R.J., Paturel, G., and Fouqu\'{e}, P. 1991,
Third Reference Catalogue of Bright Galaxies (New York: Springer-Verlag) (RC3)

\bibitem[Driver et al.\ 1995]{dri95} Driver, S.P., Windhorst, R.A., and Griffiths, R.E. 1995, \apj, 453, 48

\bibitem[Ebisuzaki et al.\ 1993]{ebi93} Ebisuzaki, T., Makino, J., Fukushige, T., Sugimoto, D., Ito, T., and Okumura, S.K.
1993, PASJ, 45, 269


\bibitem[Eggen et al.\ 1962]{egg62} Eggen, O.J., Lynden-Bell, D., and Sandage, A.R.
1962, \apj, 136, 748


\bibitem[Fall and Efstathiou 1980]{fal80} Fall, S.M., and Efstathiou, G. 1980, \mnras, 193, 189

\bibitem[Fall and Rees 1985]{fal85} Fall, S.M., and Rees, M.J. 1985, \apj, 298, 18

\bibitem[Freeman 1970]{fre70} Freeman, K.C. 1970, \apj, 160, 811


\bibitem[Gavazzi 1993]{gav93} Gavazzi, G. 1993, \apj, 419, 469



\bibitem[Giavalisco et al.\ 1996a]{gia96a} Giavalisco, M., Livio, M., Bohlin, R.C., Macchetto, F.D., and Stecher, T.P.
1996a, \aj, 112, 369


\bibitem[Giavalisco et al.\ 1996b]{gia96b} Giavalisco, M., Steidel, C.C., and Macchetto, F.D. 1996b, \apj, 470, 189


\bibitem[Glazebrook et al.\ 1995]{gla95} Glazebrook, K.,  Ellis, R., Santiago, B.,
and Griffiths, R. 1995, \mnras, 275, L19


\bibitem[Glazebrook et al.\ 1994]{gla94} Glazebrook, K., Leh\'ar, J., Ellis, R., Arag\'on-Salamanca, A.,
and Griffiths, R. 1994, \mnras, 270, L63

\bibitem[Gott and Thuan 1976]{got76}  Gott, J.R., and Thuan, T.X. 1976, \apj, 204, 649


\bibitem[Griffiths et al.\ 1994]{gri94} Griffiths, R.E. et al. 1994, \apjl, 435, L19

\bibitem[Hausman and Roberts 1984]{hau84}  Hausman.,M.A., and Roberts,W.W. 1984, \apj, 282,106

\bibitem[Henriksen 1991]{hen91} Henriksen, R.N. 1991, \apj, 377, 500

\bibitem[Holtzman et al.\ 1993]{hol93} Holtzman, J.A. et al. 1993, \aj, 106, 1826

\bibitem[Impey and Bothun 1997]{imp97} Impey, C., and Bothun, G. 1991, \araa, 35, 267

\bibitem[Infante et al.\ 1996]{inf96} Infante, L., de Mello, D.F., and Menanteau, F. 1996, \apj, 469, L85

\bibitem[Jablonka and Arimoto 1992]{jab92} Jablonka, P., and Arimoto, N. 1992, \aap, 255, 63

\bibitem[Jablonka et al.\ 1996]{jab96} Jablonka, P., Martin, P., and Arimoto, N. 1996,
  \aj, 112, 1415

\bibitem[Katz 1992]{kat92} Katz, N. 1992, \apj, 391, 502

\bibitem[Katz and Gunn 1991]{kat91} Katz, N., and Gunn, J.E. 1991, \apj, 377, 365

\bibitem[Kennicutt 1983]{ken83} Kennicutt, R.C. 1983, \apj, 272, 54

\bibitem[Kennicutt 1989]{ken89} Kennicutt, R.C. 1989, \apj, 344, 685

\bibitem[Kent 1985]{ken85} Kent, S.M. 1985, \apjs, 59, 115

\bibitem[Koo et al.\ 1996]{koo96} Koo, D.C. et al. 1996, \apj, 469, 535

\bibitem[Kormendy 1993]{kor93} Kormendy, J. 1993, in IAU Symp. 153, Galactic Bulges,
eds. H. Dejonghe and H.J. Habing (Dordrecht: Kluwer Academic Publishers), 209

\bibitem[Lacey and Fall 1985]{lac85} Lacey, C.G., and Fall, S.M. 1985, \apj, 290, 154

\bibitem[Larson 1969]{lar69} Larson, R.B. 1969, \mnras, 145, 405

\bibitem[Larson 1974]{lar74} Larson, R.B. 1974, \mnras, 166, 585

\bibitem[Larson 1975]{lar75} Larson, R.B. 1975, \mnras, 173, 671

\bibitem[Larson 1976]{lar76} Larson, R.B. 1976, \mnras, 176, 31

\bibitem[Larson and Tinsley 1978]{lar78} Larson, R.B., and Tinsley, B.M. 1978, \apj, 219, 46

\bibitem[Levinson and Roberts 1981]{lev81} Levinson,F.H., and Roberts,W.W. 1981, \apj, 245,465

\bibitem[Lin and Pringle 1987]{lin87} Lin, D.N.C., and Pringle, J.E. 1987, \mnras, 225, 607

\bibitem[Matteucci and Fran\c{c}ois 1989]{mat89} Matteucci, F., and Fran\c{c}ois, P. 
1989, \mnras, 239, 885

\bibitem[Matteucci and Brocato 1990]{mat90} Matteucci, F., and Brocato, E. 1990, \apj, 365, 539 

\bibitem[McWilliam and Rich 1994]{mcw94} McWilliam, A., and Rich, R.M. 1994, \apjs, 91, 749

\bibitem[Meier 1976]{mei76} Meier, D.L. 1976, \apj, 207, 343

\bibitem[Navarro and White 1994]{nav94} Navarro, J.F., and White, S.D.M. 1994, \mnras, 267, 401

\bibitem[Noguchi 1991]{nog91} Noguchi, M. 1991, \mnras, 251, 360

\bibitem[Noguchi 1996]{nog96} Noguchi, M. 1996, \apj, 469, 605

\bibitem[Noguchi 1998]{nog98} Noguchi, M. 1998, Nature, 392, 253

\bibitem[Norris 1996]{nor96} Norris, J.E. 1996, in IAU Symp. 169, Unsolved Problems of the Milky Way,
ed. L.Blitz and P.Teuben (Dordrecht: Kluwer Academic Publishers), 353

\bibitem[Peletier and Balcells 1996]{pel96} Peletier, R.F., and Balcells, M. 1996, \aj, 111, 2238
 
\bibitem[Persic and Salucci 1988]{per88} Persic, M., and Salucci, P. 1988, \mnras, 234, 131
 
\bibitem[Quinn and Goodman 1986]{qui86}  Quinn, P.J., and Goodman, J. 1986, \apj, 309, 472

\bibitem[Rand and Kulkarni 1990]{ran90}  Rand, R.J., and Kulkarni, S. 1990, \apj, 349, L43

\bibitem[Read et al.\ 1997]{rea97} Read, A.M., Ponman, T.J., and Strickland, D.K. 1997, \mnras, 286, 626 

\bibitem[Rich 1996]{ric96} Rich, R.M. 
1996, in IAU Symp. 169, Unsolved Problems of the Milky Way,
ed. L.Blitz and P.Teuben (Dordrecht: Kluwer Academic Publishers), 403

\bibitem[Roberts and Hausman 1984]{rob84} Roberts,W.W., and Hausman,M.A. 1984, \apj, 277,744

\bibitem[Rubin et al.\ 1980]{rub80} Rubin, V.C., Ford, W.K.,Jr., and Thonnard, N. 1980, \apj, 238, 471

\bibitem[Rubin et al.\ 1982]{rub82} Rubin, V.C., Ford, W.K.,Jr., Thonnard, N., and Burstein, D.
1982, \apj, 261, 439

\bibitem[Rubin et al.\ 1985]{rub85} Rubin, V.C., Burstein, D., Ford, W.K., Thonnard, N. 1985, \apj, 289, 81

\bibitem[Sage 1993]{sag93} Sage, L.J. 1993, \aap, 272, 123

\bibitem[Sandage 1961]{san61} Sandage, A. 1961, The Hubble Atlas of Galaxies (Washington, D.C.,
Carnegie Institution of Washington)

\bibitem[Sandage 1986]{san86} Sandage, A. 1986, \aap, 161, 89

\bibitem[Sandage et al.\ 1970]{san70} Sandage, A.R., Freeman, K.C., and Stokes, N.R. 1970, \apj, 160, 831

\bibitem[Schombert et al.\ 1992]{sch92} Schombert, J.M.,
Bothun, G.D., Schneider, S.E., and McGaugh, S.S. 1992, \aj, 103, 1107

\bibitem[Shlosman and Noguchi 1993]{shl93} Shlosman, I., and Noguchi, M. 1993, \apj, 414, 474

\bibitem[Simien and de Vaucouleurs 1986]{sim86} Simien, F., and de Vaucouleurs, G. 1986, \apj, 302, 564

\bibitem[Solomon et al.\ 1979]{sol79} Solomon, P.M., Sanders, D.B., and Scoville, N.Z. 1979, in IAU Symp. 84,
The Large-Scale Characteristics of the Galaxy, ed. W.B.Burton (Dordrecht:
Reidel), 35

\bibitem[Sprayberry et al.\ 1995]{spr95} Sprayberry, D., Impey, C.D., Bothun, G.D., and Irwin, M.J.
1995, \aj, 109, 558

\bibitem[Stark and Brand 1989]{sta89} Stark, A.A., and Brand, J. 1989, \apj, 339, 763

\bibitem[Steidel 1990]{ste90} Steidel, C.C. 1990, \apjs, 74, 37

\bibitem[Steidel et al.\ 1996a]{ste96a} Steidel, C.C., Giavalisco, M., Dickinson, M., and Adelberger, K.L.
1996a, \aj, 112, 352

\bibitem[Steidel et al.\ 1996b]{ste96b} Steidel, C.C., Giavalisco, M., Pettini, M., Dickinson, M., and Adelberger, K.L.
1996b, 
\apj, 462, L17

\bibitem[Steinmetz and M\"uller 1995]{ste95} Steinmetz, M., and M\"uller, E. 1995, \mnras, 276, 549


\bibitem[Tinsley 1981]{tin81} Tinsley, B.M. 1981, \mnras, 194, 63

\bibitem[T\'{o}th and Ostriker 1992]{tot92}
T\'oth, G., and Ostriker, J.P. 1992, \apj, 389, 5

\bibitem[Toomre 1964]{too64} Toomre, A. 1964, \apj, 139, 1217

\bibitem[Tully et al.\ 1982]{tul82} Tully, R.B., Mould, J.R., and Aaronson, M. 1982, \apj, 257, 527

\bibitem[van den Bergh 1976]{vdb76} van den Bergh, S. 1976, \apj, 206, 883

\bibitem[van den Bergh et al.\ 1996]{van96} van den Bergh, S., Abraham, R.G., Ellis, R.S., Tanvir, N.R., Santiago, B.X.,
and Glazebrook, K.G. 1996, \aj, 112, 359

\bibitem[van der Kruit and Searle 1982]{van82} van der Kruit, P.C., and Searle, L. 1982, \aap, 110, 61

\bibitem[van der Kruit and Shostak 1984]{van84} van der Kruit, P.C., and Shostak, G.S. 1984, \aap,
134, 258

\bibitem[Vazdekis et al.\ 1996]{vaz96} Vazdekis, A., Casuso, E., Peletier, R.F., and Beckman, J.E.
1996, \apjs, 106, 307

\bibitem[Whitmore 1984]{whi84} Whitmore, B.C. 1984, \apj, 278, 61

\bibitem[Yoshii and Arimoto 1987]{yos87} Yoshii, Y., and Arimoto, N. 1987, \aap, 188, 13

\bibitem[Yoshizawa and Wakamatsu 1975]{yos75} Yoshizawa, M., and Wakamatsu, K. 1975, \aap, 44, 363

\bibitem[Young 1990]{you90} Young, J.S. 1990, in 
The Interstellar Medium in Galaxies, eds. H.A.Thronson, \& J.M.Shull
(Dordrecht: Kluwer Academic Publishers), 67

\bibitem[Young et al.\ 1989]{you89} Young, J.S., Xie, S., Kenney, J.D.P., and Rice, W.L.
1989, \apjs, 70, 699

\end{thebibliography}
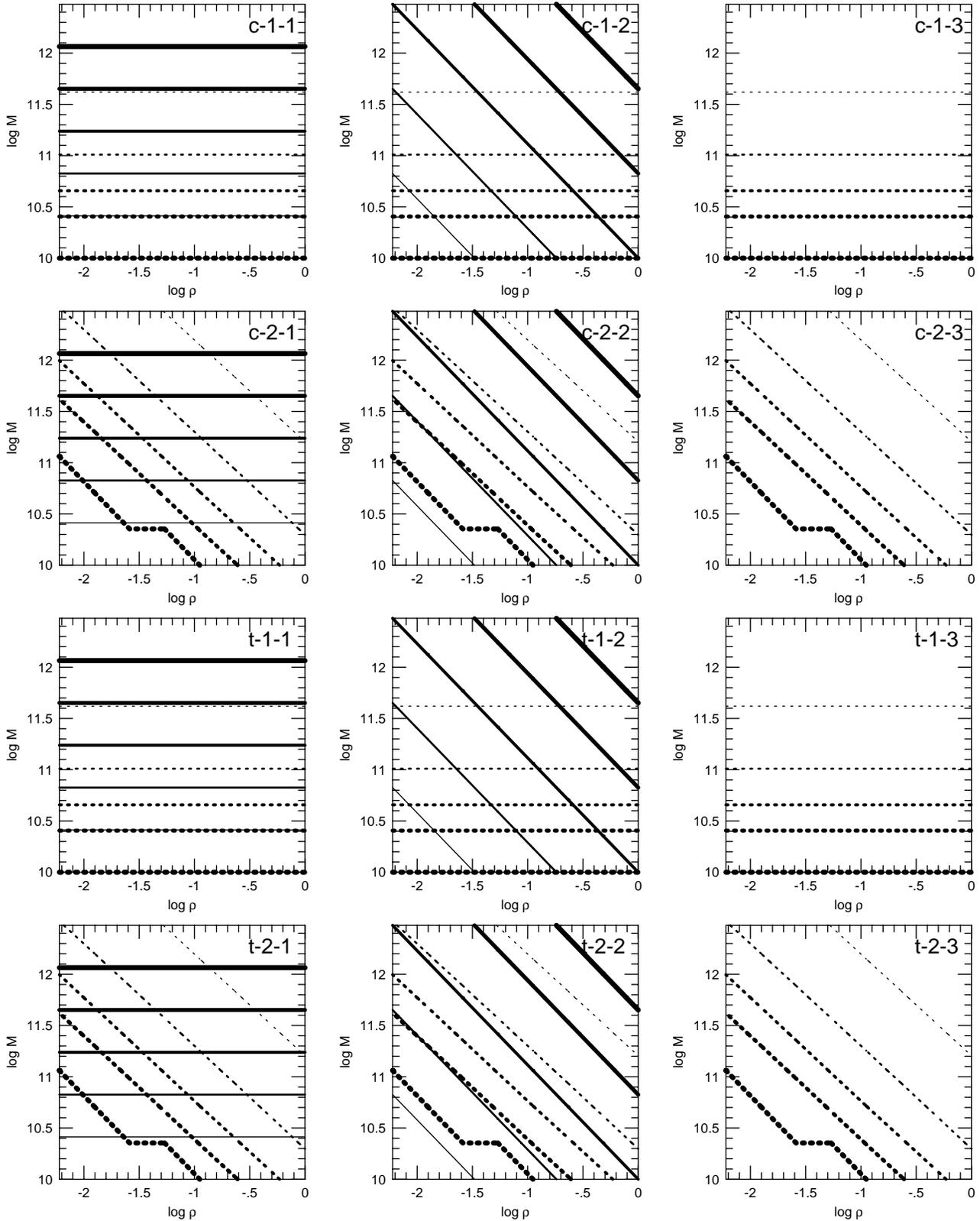
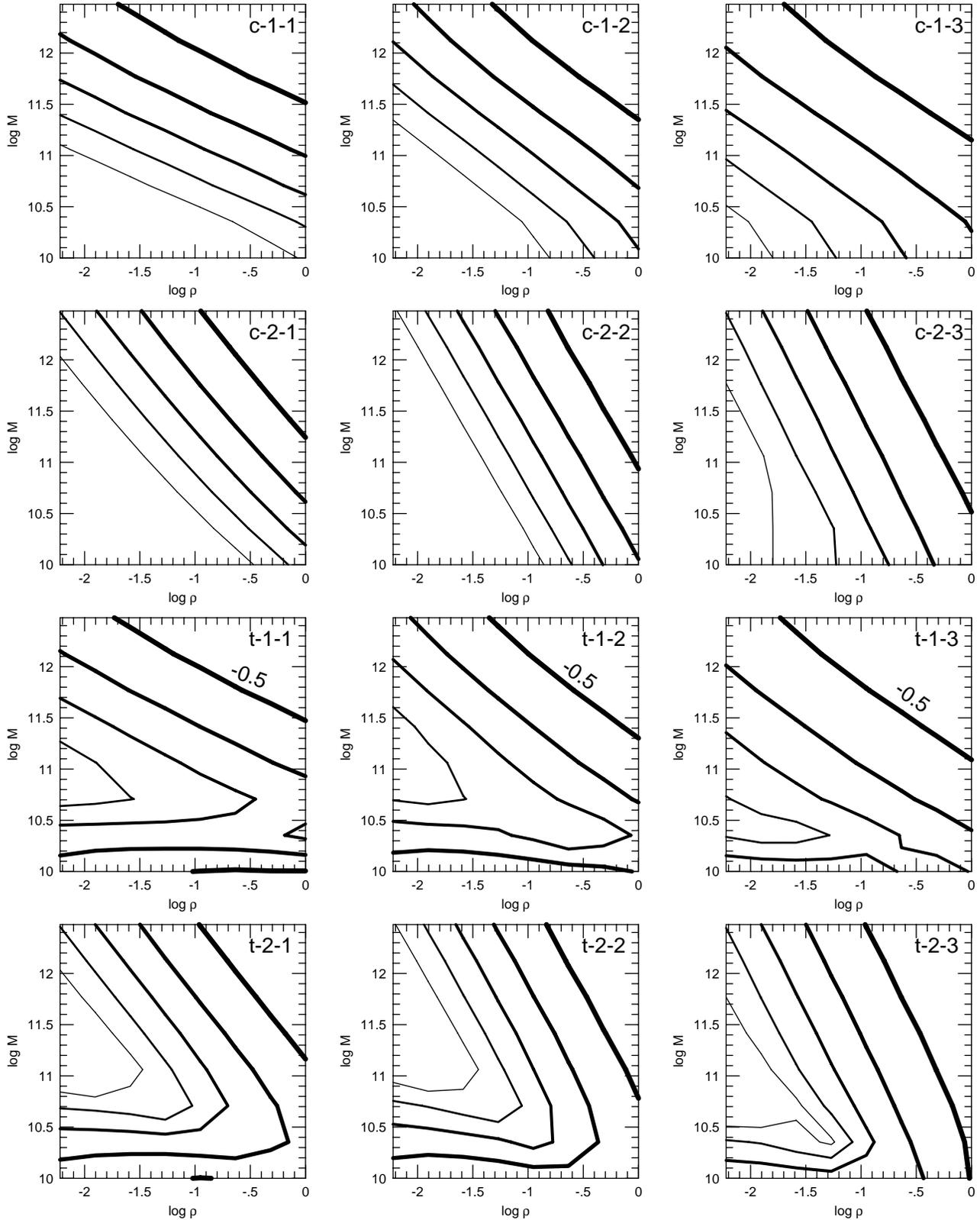
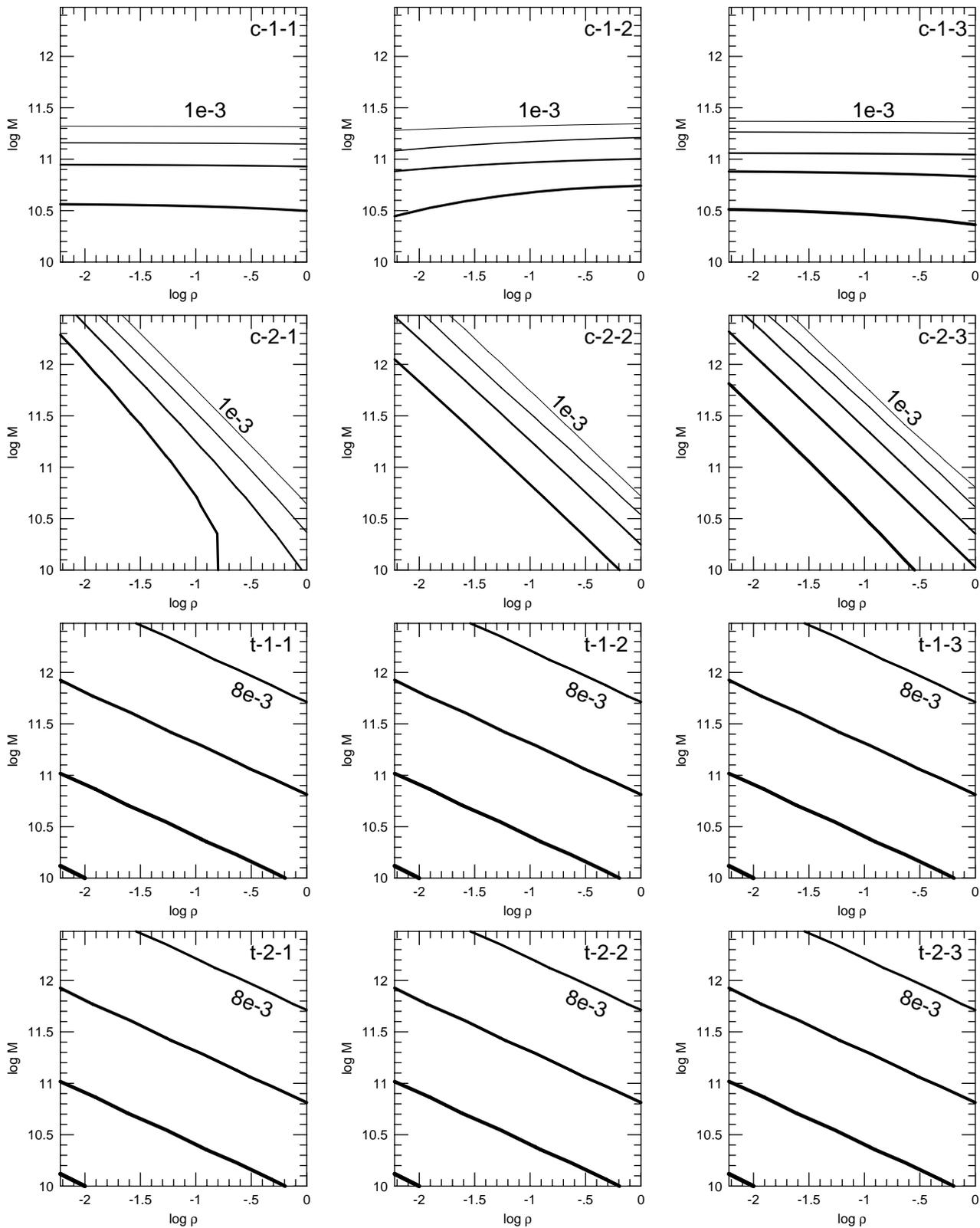
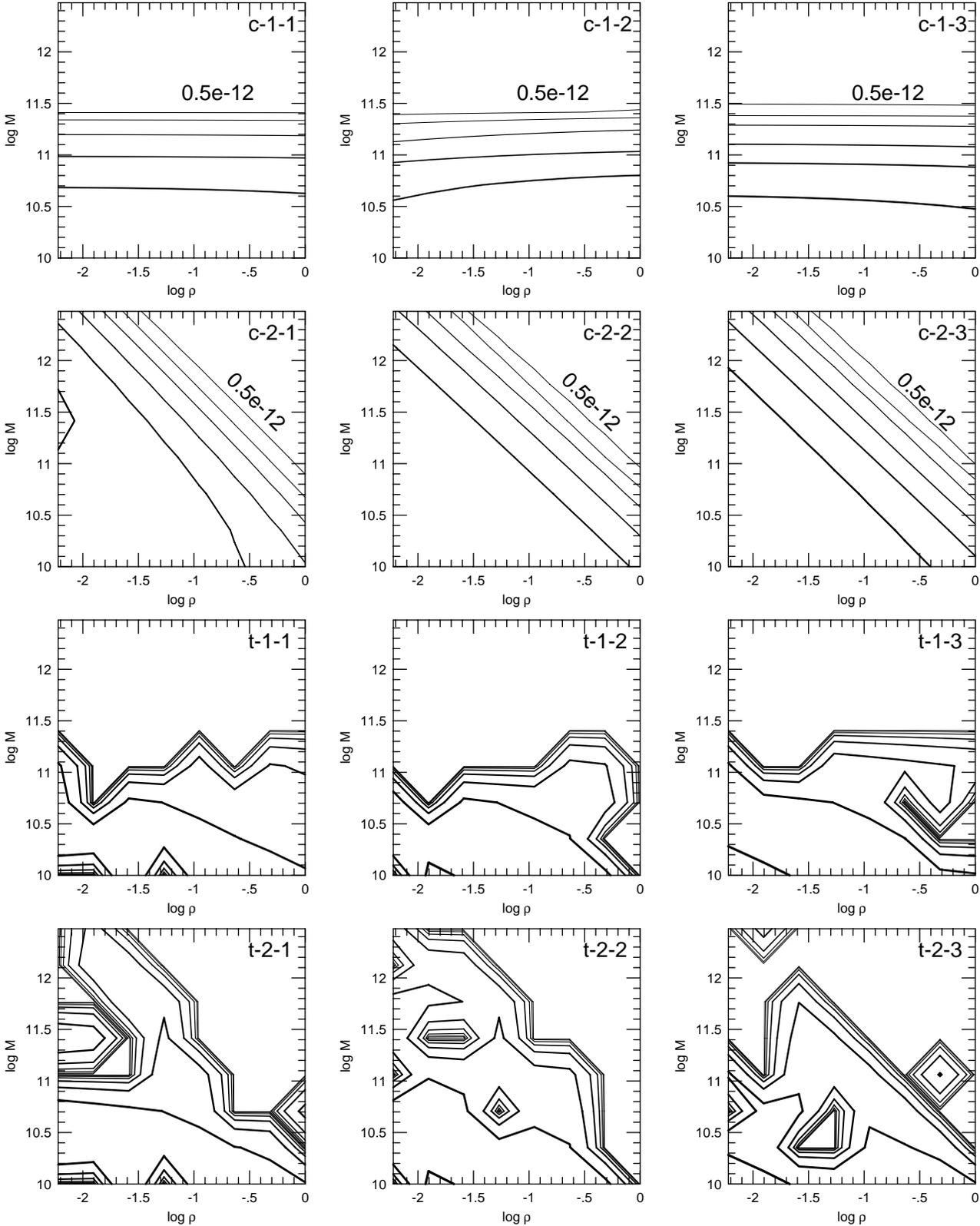
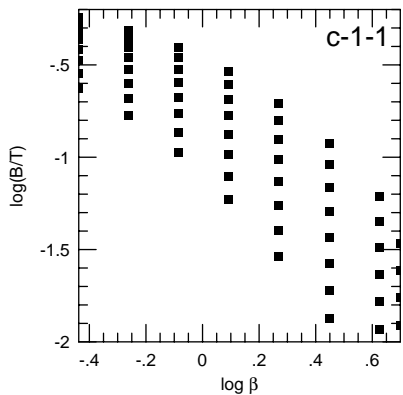
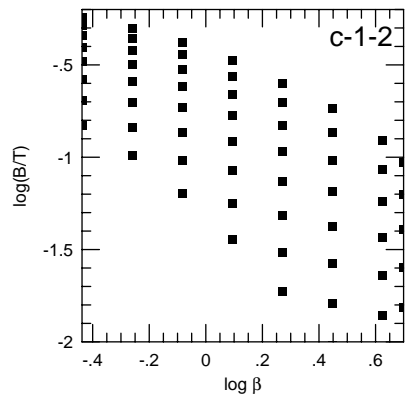
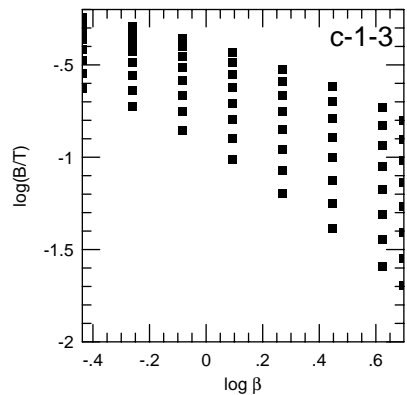
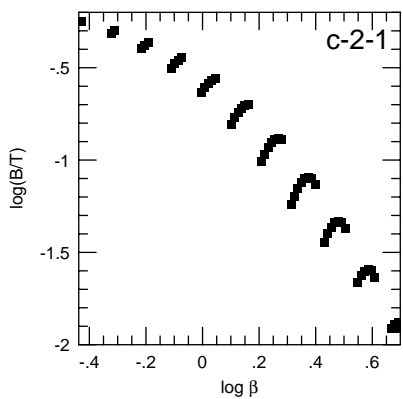
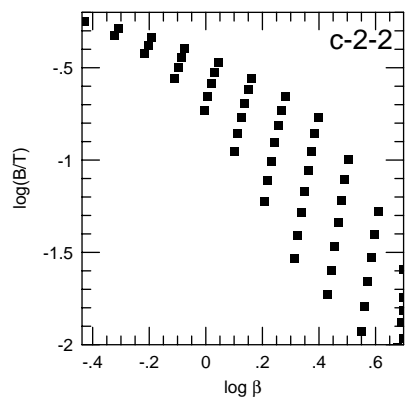
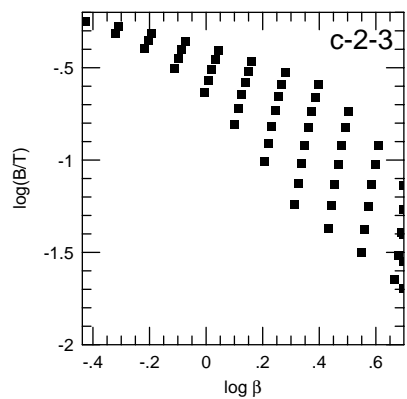
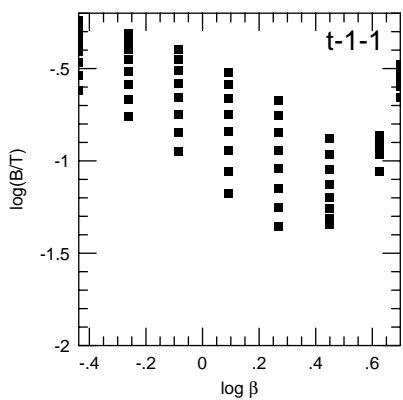
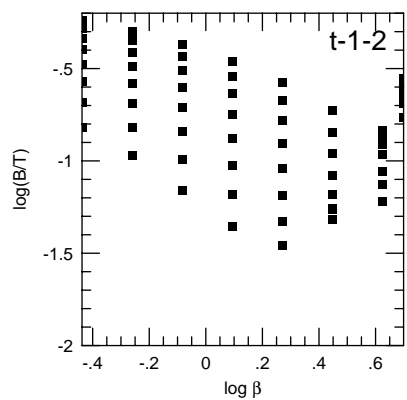
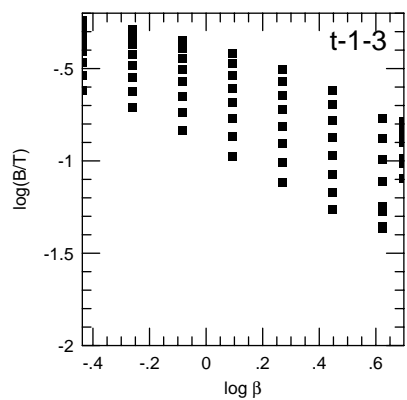
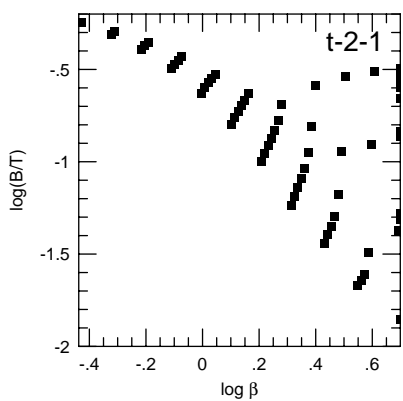
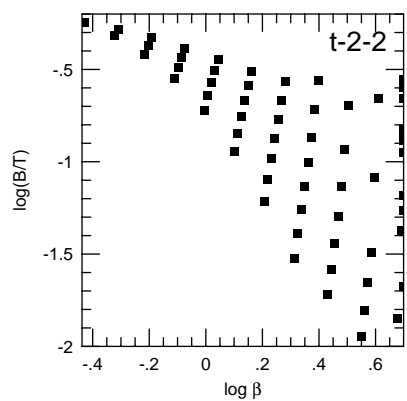
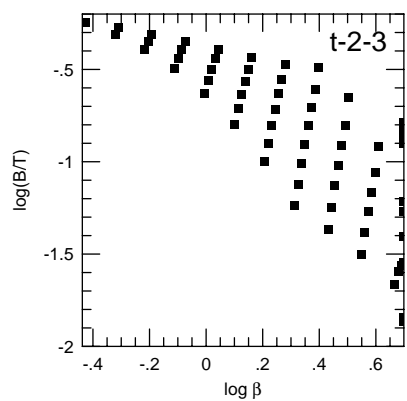
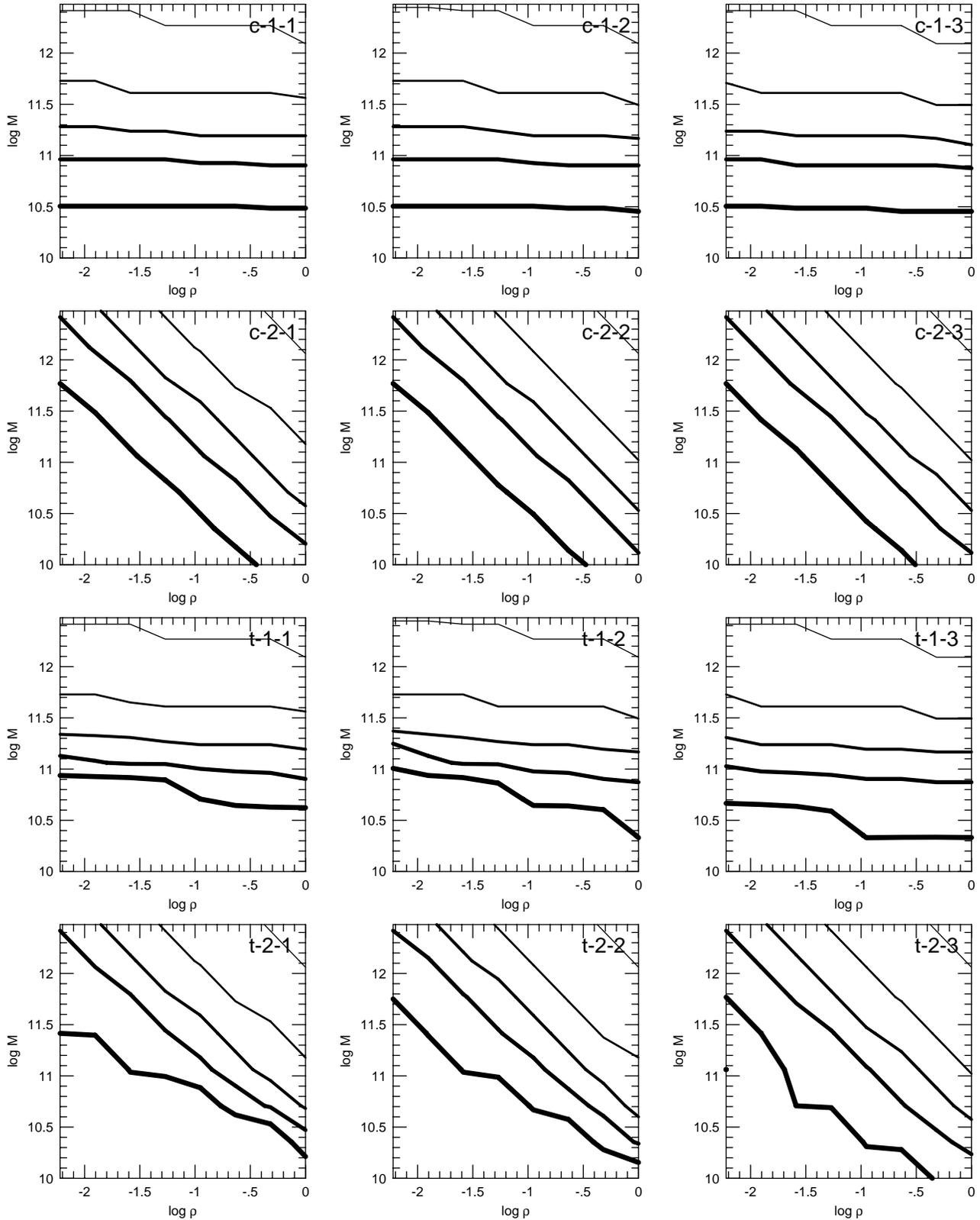
\end{document}